\documentclass{emulateapj}
\usepackage{apjfonts}
\usepackage{graphics,graphicx,rotating}
\usepackage{natbib}
\usepackage{xspace}
\usepackage{amssymb}
\usepackage{amsmath}
\usepackage{epstopdf}
\usepackage{subfigure}

\usepackage{color}
\usepackage{soul}

\shorttitle{The Pre-merger Impact Velocity of the Binary Cluster A1750}
\shortauthors{Molnar, et al.}

\newcommand{\simless} 
     {\ensuremath{\lower 3pt\hbox{$\rlap{\raise5pt\hbox{$\char'074$}}\mathchar"7218$}}}
\newcommand{\simgreat}
     {\ensuremath{\lower 3pt\hbox{$\rlap{\raise5pt\hbox{$\char'076$}}\mathchar"7218$}}}

\newcommand{\simgt}{\lower.5ex\hbox{$\; \buildrel > \over \sim \;$}}
\newcommand{\simlt}{\lower.5ex\hbox{$\; \buildrel < \over \sim \;$}}

\newcommand{\nop}{{\noindent}}

\newcommand{\LCDM}{{\sc $\Lambda$CDM}}

\newcommand{\CHANDRA}{{\textit{Chandra}}}
\newcommand{\APEC}{{\textit{APEC}}}

\newcommand{\XMM}{{\textit{XMM-Newton}}}
\newcommand{\ROSAT}{{\textit{ROSAT}}}
\newcommand{\ASCA}{{\textit{ASCA}}}

\newcommand{\FLASH}{{\sc FLASH}}
\newcommand{\HE}{hydrostatic equilibrium}

\newcommand{\DEGREE}{{$\!\!^{\circ}$}}
\newcommand{\ASEC}{\ensuremath{\arcsec}}
\newcommand{\AMIN}{\ensuremath{\arcmin}}

\newcommand{\TMSUNFOUR}{{$\times 10^{\,14}\,$\ensuremath{\mbox{\rm M}_{\odot}}}}

\newcommand{\KMSEC}{{$\rm km\;s^{-1}$}}

\newcommand{\rmsub}[1]{\ensuremath{_{\rm #1}}}
\newcommand{\RVIR}{\ensuremath{R\rmsub{vir}}}

\newcommand*{\ltsim}{\ {\raise-.75ex\hbox{$\buildrel<\over\sim$}}\ }
\newcommand*{\gtsim}{\ {\raise-.75ex\hbox{$\buildrel>\over\sim$}}\ }
\newcommand*{\proptosim}{\ {\raise-.75ex\hbox{$\buildrel\propto\over\sim$}}\ }

\newcommand{\CLJMAS}{{\sc CL J0152-1347}}
\newcommand{\MACSJSEVEN}{{\sc MACS J0744.8+3927}}
\newcommand{\CLJONETWO}{{\sc CL J1226.9+3332}}
\newcommand{\MACSJOSEVEN}{{\sc MACS J0717.5+3745}}

\newcommand{\SMALLER}{{\;\!}}

\begin{document}

\title{The Pre-merger Impact Velocity of the Binary Cluster A1750 
         from X--ray, Lensing and Hydrodynamical Simulations
}

\author{
Sandor. M. Molnar\altaffilmark{1}, I-Non Chiu\altaffilmark{2}, Tom Broadhurst\altaffilmark{3,4}, 
and 
Joachim G. Stadel\altaffilmark{5}, 
}

\altaffiltext{1}{Leung Center for Cosmology and Particle Astrophysics, 
                      National Taiwan University, Taipei 10617, Taiwan, sandor@phys.ntu.edu.tw}

\altaffiltext{2}{Department of Physics, Ludwig-Maximilians University, Scheinerstr  1, Munich, Germany}

\altaffiltext{3}{Department of Theoretical Physics, University of the Basque Country, Bilbao 48080, Spain}
                                            
\altaffiltext{4}{Ikerbasque, Basque Foundation for Science, Alameda Urquijo, 36-5 Plaza Bizkaia 48011, Bilbao, Spain}
                  
\altaffiltext{5}{Institute for Theoretical Physics, University of Zurich, 8057 Zurich, Switzerland}

\begin{abstract}
Since the discovery of the ``bullet cluster'' several similar cases
have been uncovered suggesting relative velocities well beyond 
the tail of high speed collisions predicted by the concordance \LCDM\ model. 
However, quantifying such post--merger events with
hydrodynamical models requires a wide coverage of possible initial conditions. 
Here we show that it is simpler to interpret pre--merger cases, such as A1750, where the gas 
between the colliding clusters is modestly affected, so that the initial conditions are clear.  
We analyze publicly available \CHANDRA\ data confirming a significant increase in 
the projected X--ray temperature between the two cluster centers in A1750 
consistent with our expectations for a merging cluster.
We model this system with a self-consistent hydrodynamical simulation of dark 
matter and gas using the \FLASH\ code.
Our simulations reproduce well the X-ray data, and the measured
redshift difference between the two clusters in the phase before the first core passage 
viewed at an intermediate projection angle. 
The deprojected initial relative velocity derived using our model is 1460 \KMSEC\
which is considerably higher than the predicted mean impact velocity for simulated massive 
haloes derived by recent \LCDM\ cosmological simulations, but it is within the allowed range.
Our simulations demonstrate that such systems can be identified using a
multi--wavelength approach and numerical simulations, for which the 
statistical distribution of relative impact velocities may provide
a definitive examination of a broad range of dark matter scenarios.
\end{abstract}


\section{Introduction}
\label{S:Intro}

Direct evidence of the existence of dark matter can be seen in the
case of the ``bullet cluster'' (CL0152-1357, \citealt{Markevitch2004ApJ606}), 
where the infalling cluster gas displays a bow shock followed by a wedge shaped
contact discontinuity generated by the infalling gas.  It has been
found that the center of the mass belonging to the infalling cluster
is offset from its gas, providing direct evidence of the existence of
a dark mass component (\citealt{Clowe04}; see also \citealt{Clowe06,Bradac06}).  
Numerical simulations of binary galaxy cluster mergers show that the bullet cluster 
should have had an initially high impact velocity \simgreat\ 3000 \KMSEC\ 
\citep{SpriFarr2007MNRAS380p911,MastBurk08MNRAS389p967}.

A series of estimates have been made to quantify the probability of
finding a system with a large impact velocity implied by the bullet
cluster using large scale cosmological N--body simulations
\citep{Lee10,Hayashi06}.  The conclusion was that it is very unlikely
that merging clusters have very high impact velocities in the context
of \LCDM.  However, it was pointed out by \cite{Lee10} that the
statistic of high impact velocity mergers is on the steeply falling
tail of the probability distribution, therefore their conclusion is 
sensitive to model uncertainties.  Recently, \cite{ThomNaga2012MNRAS419}
carried out large scale cosmological simulations to estimate the
effect of the box size of the simulations on the high velocity tail of
the probability distribution.  Thompson \& Nagamine found that the
probability of finding one cluster merger with impact velocity 
\simgreat\ 3000 \KMSEC\ in a concordance \LCDM\ model is $3 \times
10^{-8}$, i.e., it is very unlikely (the peak of the impact velocity
probability distribution is at about 550 \KMSEC, see their Figure 15).
They concluded that the bullet cluster is either incompatible with the
concordance \LCDM\ model, or the initial conditions suggested by the
non-cosmological binary merger simulations overestimate the 
relative impact velocity.

Since the discovery of the bullet cluster several other merging clusters
have been found with large offsets between dark matter and X--ray gas
emission, as well as between peaks of X--ray emission and SZ signal
(\CLJMAS: \citealt{Massardi2010APJL718},
\MACSJSEVEN\ and \CLJONETWO: \citealt{Korngut2011ApJ734}, 
\MACSJOSEVEN: \citealt{Mroczkowski2012APJ}).
The size of these offsets is determined by two opposing interactions:
the attractive force of the massive dark matter halo on its own gas
and the ram pressure exerted on the gas of the infalling cluster by
the gas of the more massive interacting cluster.  
Using self--consistent N/body--hydrodynamical simulations of merging galaxy
clusters, \cite{Molnet2012ApJ748} demonstrated that, in general, a
large impact velocity is necessary to explain 100 kpc scale offsets
between the peaks of mass distribution, X--ray emission and SZ signal.

Clearly, a confirmation that several cluster mergers have very large
impact velocities would pose a serious problem for our concordance
\LCDM\ model, so it is important to explore fully shortcomings of 
the predictions in this context.  Typically, binary galaxy cluster
simulations are semi--adiabatic in the treatment of gas physics, in the
sense that they contain only adiabatic processes plus shock heating,
but they do not contain non-gravitational physics such as feedback
mechanisms, cooling and heating, nor magnetic fields. However, these
effects might be minor relative to the very energetic shocks galaxy
mergers produce. It is also possible that the basic non-interacting
particle premise for dark matter is not the best assumption. 
A possible alternative is a very light and very cold bosonic dark matter
(less than about 10$^{-24}$ eV and a current temperature of a fraction
of 1 K) so that the dark matter would then reside in a Bose--Einstein condensate.
It has been shown, that axions may form Bose--Einsten
condensates, which make them a viable candidate for this type of dark matter
\citep{SikiYanf2009PhRvL103}. Cosmological simulations of the growth of structure
in this state show interesting macroscopic interference modifying the
dynamics of halo collisions \citep{WooChiueh2009ApJ697} with possibly 
interesting implications for the interpretation of colliding clusters.

Bose-Einstein condensate in a cosmological context have been studied
recently by several authors using the Gross--Pitaevskii equation
(\citealt{HARKO11PhRvD83,MadTot12,Kain2012PhRvD85}, and references therein). 
On large scales, after photon decoupling, the BAC dark matter behaves similarly 
to conventional dark matter, therefore structure formation occurs similarly to the
concordance \LCDM\ (assuming that structure formation via CDM and
BACDM result similar initial conditions after photon decoupling,
however, see \citealt{HARKO2011MNRAS.413}).  There are some
differences in expansion rates and density contrasts \citep{Kain2012PhRvD85}.  
However, structure formation will differ qualitatively in the
highly non--linear regime due to quantum mechanical effects associated
with the Bose--Einstein condensate dark matter.

In this paper we interpret multi-wavelength observations of A1750 with
self--consistent N--body/hydrodynamical numerical simulations using a
parallel adaptive mesh code, \FLASH, to constrain the initial impact
velocity in this merging system.  We limit the range of input masses
for the pair using weak lensing results from previous Subaru observations of
A1750, the observed relative velocities of each component derived
from member galaxy spectroscopy, and the projected temperature and
X--ray morphology that we derive from publicly available (\CHANDRA)
observations.

The overall structure of this paper is as follows.  We summarize
previous results on A1750 based on X--ray, and Optical/Infrared (O/IR)
observations in section \ref{S:A1750}.  We describe our X--ray 
data analysis in Section \ref{S:Data}. In Section \ref{S:FLASH},
we describe our \FLASH\ simulations, and our methods to obtain
simulated X--ray images and projected temperature maps.  We summarize
our results and discuss how our method can be used to interpret
multi--wavelength observations of merging clusters and determine the
impact velocities in Section~\ref{S:Results}. 
Unless stated otherwise, errors and limits quoted on physical quantities are 68\% CL.

\section{Abell 1750}
\label{S:A1750}

A1750 comprises two massive cluster components located at a mean
redshift of $z$ = 0.086.  It was studied preciously using Einstein,
\ROSAT, \ASCA, and \XMM, X-ray satellites
\citep{Formet81ApJ243L,Donnet01ApJ562,Belset2004AA415} 
and with weak lensing applied to deep Subaru imaging by \cite{Okab2008PASJ60}.

Deep \XMM\ X-ray data of A1750 was analyzed in detail by
\cite{Belset2004AA415}, finding an X-ray brighter cluster at the
 center of the field of view, and a fainter cluster to the North,
 termed A1750C and A1750N respectively, separated by about $R_p 900$ kpc, in
 projection on the sky.  The X--ray emission at the center of A1750C is
 elongated towards A1750N, and more circular farther from the center.
 The cD galaxy of A1750C is offset from the X-ray peak towards the
 East, which is also the direction the gas seems to be compressed
 (as suggested by the X--ray morphology).
 Belsole et al. concluded that A1750C is not a well relaxed cluster as there is 
 no evidence for a cool core in A1750C, but instead an excess of entropy. 
 They found that A1750N is relatively more relaxed, with an approximately 
 circular X-ray morphology, a cool core and the center of the X-ray emission
 seems to coincide with the two bright central galaxies.  The X-ray emission
 seems to be extended towards A1750C, which we expect for merging
 clusters in an early stage.

The global spectrum for the two cluster components and the interaction region
in between them are found to have projected temperatures:
3.87$\pm$0.10 keV, 2.84$\pm$0.12 keV and 5.12$^{+0.77}_{-0.69}$ keV
with 90\% CL \citep{Belset2004AA415}.  
The X-ray temperatures of about 3 and 4 keV of the cluster components suggest 
that the sum of the virial radii of the two clusters are much larger than 1 Mpc.  
The mass--temperature and mass--virial radius scalings suggest that the two clusters 
are at a distance of about 2$\times$R$_{500}$, i.e, their intracluster gas
components are already significantly interacting.  The temperature of about
5 keV of the interaction region is much higher than the temperature we
would expect at about R$_{500}$ from the center of these clusters,
therefore it seems to confirm that these two main components are in
the early process of merging.  Based on a rough idealized model,
\cite{Belset2004AA415} estimated a Mach number of 1.64 for the merger
shocks by applying the Rankine--Hugoniot jump conditions.

Analyzing Subaru data, \cite{Okab2008PASJ60} found no significant
offset in A1750 between the position of the peaks of mass, X--ray
emission peaks, and smoothed optical luminosity.  They found no
significant mass substructure in A1750 either, thus they concluded
that most likely this cluster has no recent major merger, and hence
favor a premerger configuration.

%
%
\begin{figure}
     \centering
     \subfigure{
          \includegraphics[width=.43\textwidth]{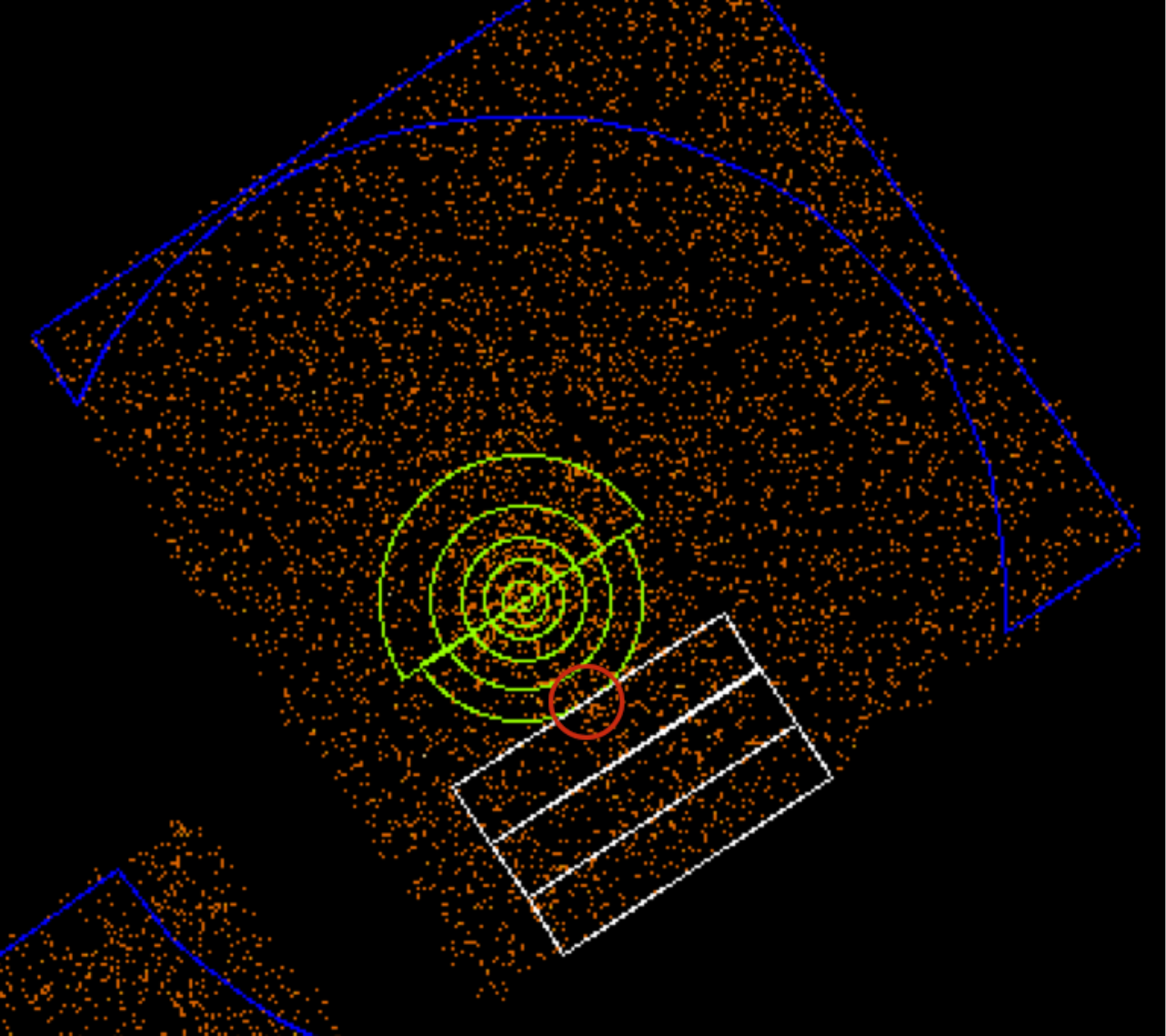}}
     \subfigure{
          \includegraphics[width=.43\textwidth]{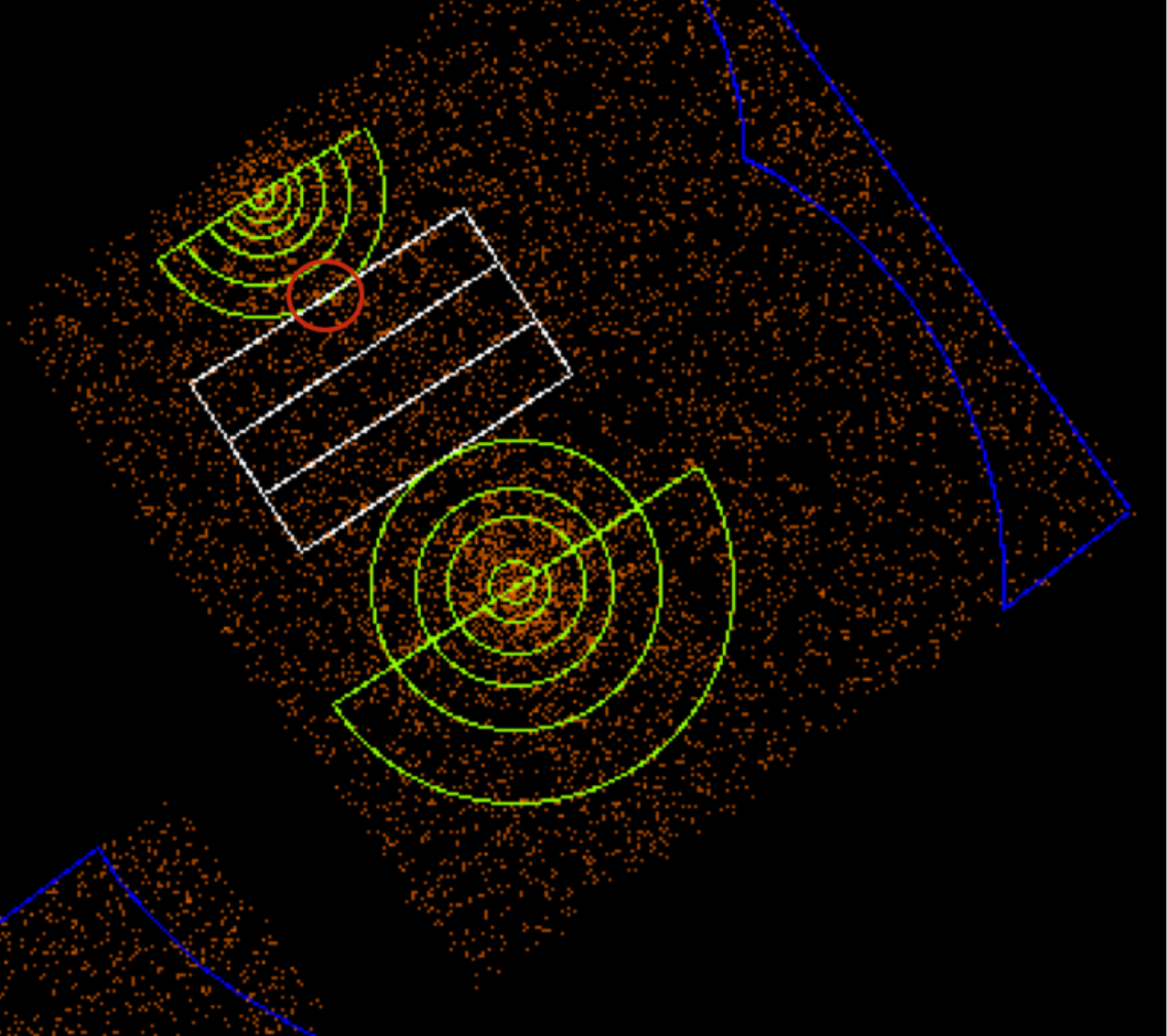}}
\caption{ \protect
 Extraction regions for spectra in A1750. Observation IDs: 11878 and 11879 (upper
 and lower image). 
 Green semi--annuli around the two cluster centers show the extraction regions for
 cluster emission, white rectangles show the regions we used for the interaction area. 
 The red circles represent regions with substructure which we excised from our analysis.
 Regions excluded due to point source contamination are not shown to avoid clutter.
 The extraction regions for the background are represented with blue lines. 
\medskip
\label{F:CHANDRA}
}
\end{figure}

\smallskip
\section{CHANDRA X--ray Data Analysis}
\label{S:Data}

In our X-ray analysis, we make use of two sets of observations of
A1750 from the \CHANDRA\ Data Archive (CDA) with ObsIDs: 11878 and
11879, and total exposure times of 19.77 ks and 20.05 ks respectively.
We follow the standard data
processing\footnote{$http://cxc.harvard.edu/ciao4.4/guides/acis\_data.html$}
to reproduce event two files (evt2) from the event one files (evt1)
using the latest {\it Chandra Interactive Analysis of Observation} (CIAO
4.4.1) with the updated {\it Calibration Data Base} (CALDB 4.5.1).  Using
the ev2 files, we detected and excluded the point sources using the
CIAO script \textit{celldetect}.  Only a small fraction of the pixels
were removed in the region we are interested in.  The strong background
flares, which are usually attributed to the high energy cosmic rays
and would increase the event rates substantially higher, were cleaned
by correcting the Good Time Interval (GTI) using the script
\textit{lc\_clean}, excluding observation times for which the event
rates were above or below 3$\sigma$.  After all corrections had been
applied, the total net exposure times were 11.515 ks and 12.481 ks for
ObsIDs 11878 and 11879.

When analyzing targets with low X-ray surface brightness spacial care
is needed to determine the background spectrum.  Although most of
the high energy particle background (e.g. cosmic rays) can be detected
and filtered out during data processing, there is still some
background contamination remaining.  This background has two
main components, the diffuse soft X-ray continuum background which
dominates the energy below about 2--3 keV, and the higher energy cosmic
rays randomly hitting the detectors.  We excluded observations in time
intervals in which they could be contaminated by cosmic rays using the
standard method of running the script \textit{lc\_clean} with
3$\sigma$-clipping.  The remaining X-ray background is customarily
removed using background subtraction.  The \CHANDRA\ science team
provides the blank-sky observations for the soft band background
estimation, however, the blank-sky background observations were
performed using the FAINT mode, while the two observations of A1750
the VFAINT mode was used.  Also, the X-ray background shows
fluctuations in different parts of the sky, therefore we choose to use
local backgrounds.  Thanks to the 16\SMALLER\AMIN$\times$16\SMALLER\AMIN\ field of
view of ACIS-I, A1750 fell only on two ACIS-I chips, we were able to
extract spectra locally, from the other two ACIS-I chips and another
front illuminated chip (ccd$\_$id = 7). This assures us that the
background extraction regions are far enough from the peak emission
($\ge$ 11\AMIN) where the cluster emission is negligible.  We also
compared the resulting fitted temperature profiles of A1750 using
local background spectra and the blank--sky spectra normalized in the
9.0--12.0 keV band, and found that they were consistent.
  
In order to study the temperature along the merging axis, we extracted
two sets of semi--annuli centered on the peaks of A1750N and A1750C,
[RA; DEC]=[13:31:10.94; $-1$:43:41.65] and [13:30:49.88;
$-$1:51:46.7]. The merging axis goes through the middle of the
semi--annuli as it is shown in Figure~\ref{F:CHANDRA}.  We found a
region with highly scattered temperature distribution (with
temperatures above 12 keV) around the peak of A1750N at 
RA = 13:31:05.342, DEC = 01:45:45.78 due to a substructure 
(\citealt{Okabet2011ApJ741}).  
We excluded this substructure from
our analysis using a circular region with a radius of 45\SMALLER\ASEC\
around the peak of A1750N.  We extracted spectra for the interacting
region between the two peaks in three rectangles with length of
400\SMALLER\ASEC\ and width of 82.6\SMALLER\ASEC\ with their short symmetry axis
aligned with the line of merging (see Figure~\ref{F:CHANDRA}).

We used the CIAO-4.4 software tool \textit{specextract} to create the
spectrum, the Response Matrix File (RMF) and the Ancillary Response
File (ARF) for each region.  The background spectra were extracted
using the CIAO-4.4 software tool \textit{dmextract} independently with
the source spectra.  We use the X-ray fitting package
\textsf{XSPEC-12.5} in our spectral analysis.  The gas emission in
each annulus is described by an optically thin plasma emission model,
\APEC, multiplied by the photo-electric absorption model,
\textit{WABS}.  We fixed the galactic photoelectric column density
absorption\footnote{from http://heasarc.gsfc.nasa.gov/docs/tools.html}
at 2.37$\times 10^{20}$ cm$^{-2}$ and the redshift at 0.086.  We
fitted the projected temperature, abundance and normalization
parameters for each semi--annulus in the energy band 0.5--7.0 keV
using C-statistic.

%
%
\begin{figure}
\centerline{
\includegraphics[width=.5\textwidth]{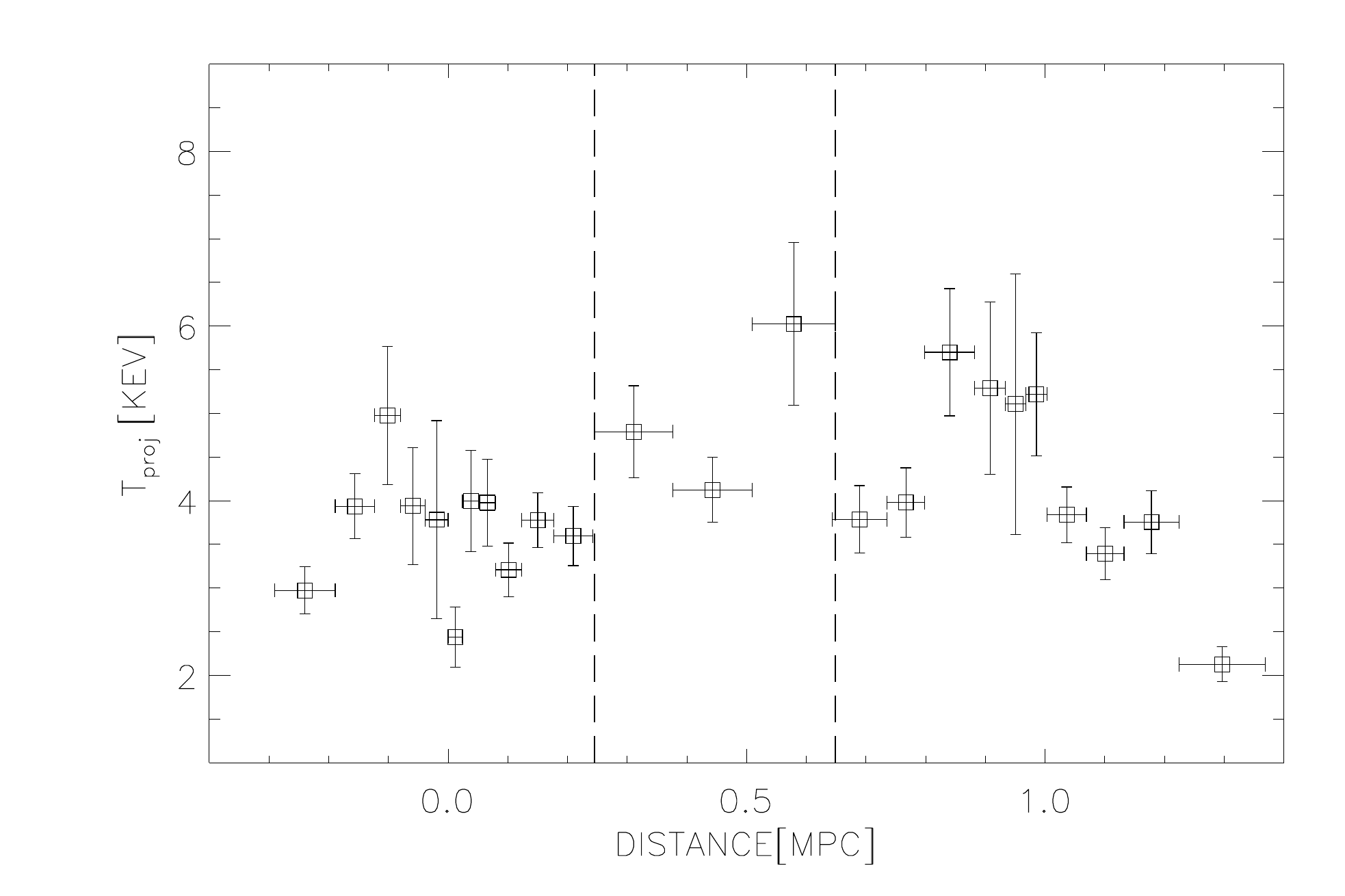}
}
\caption{\protect
Projected temperature in A1750 from \CHANDRA\ observations (points with error bars).
The gray vertical dashed lines show the centers of the two components A1750N and A1750C. 
\medskip
\label{F:TSPEC_OBS}
}
\end{figure} 

%
%
\begin{figure*}
\centering{
   \subfigure{
          \includegraphics[width=.25\textwidth]{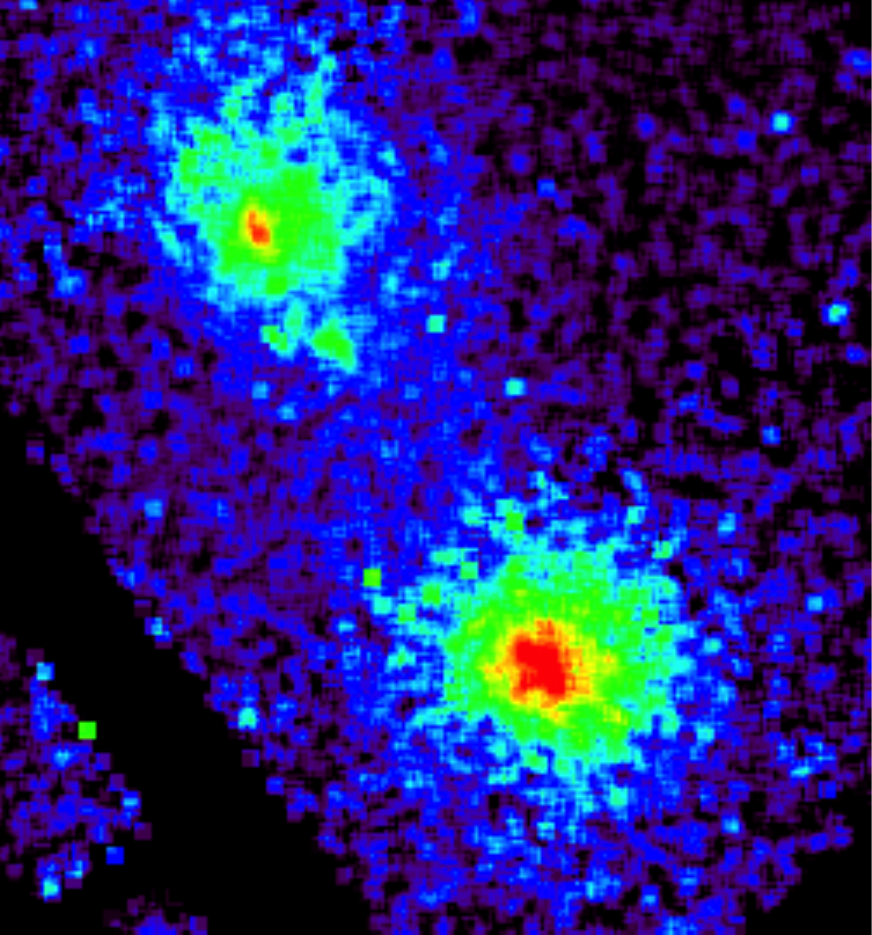}
          \includegraphics[width=.26\textwidth]{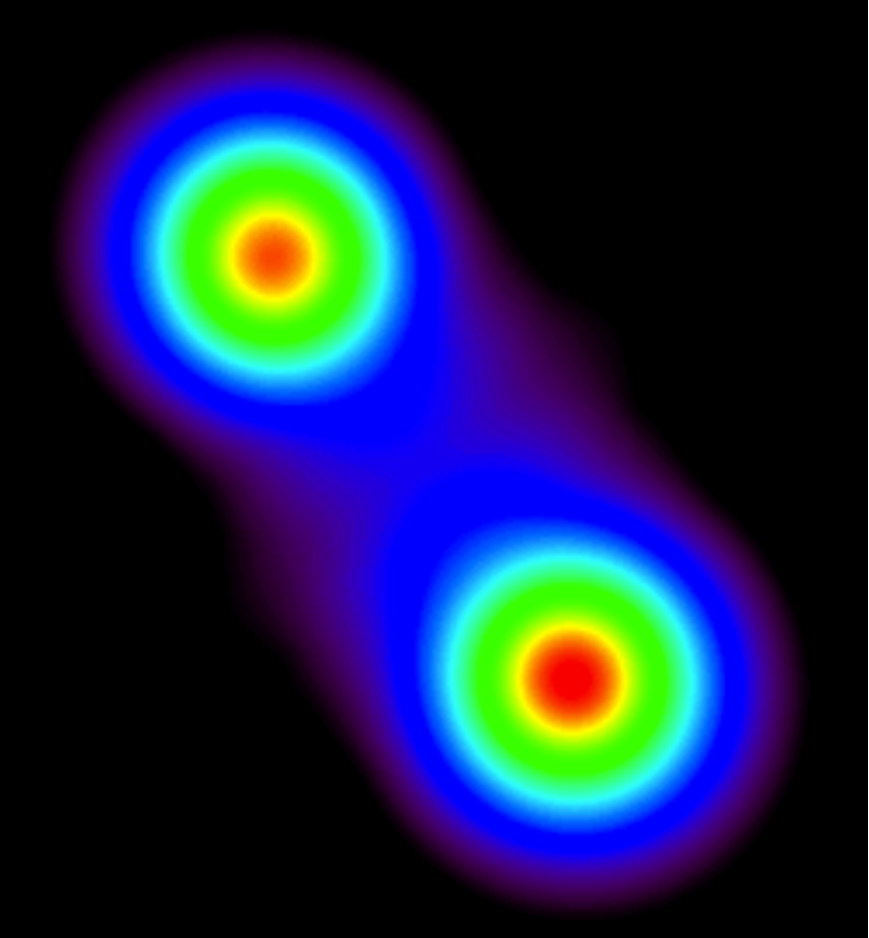}
   }
   \subfigure{
          \includegraphics[width=.28\textwidth]{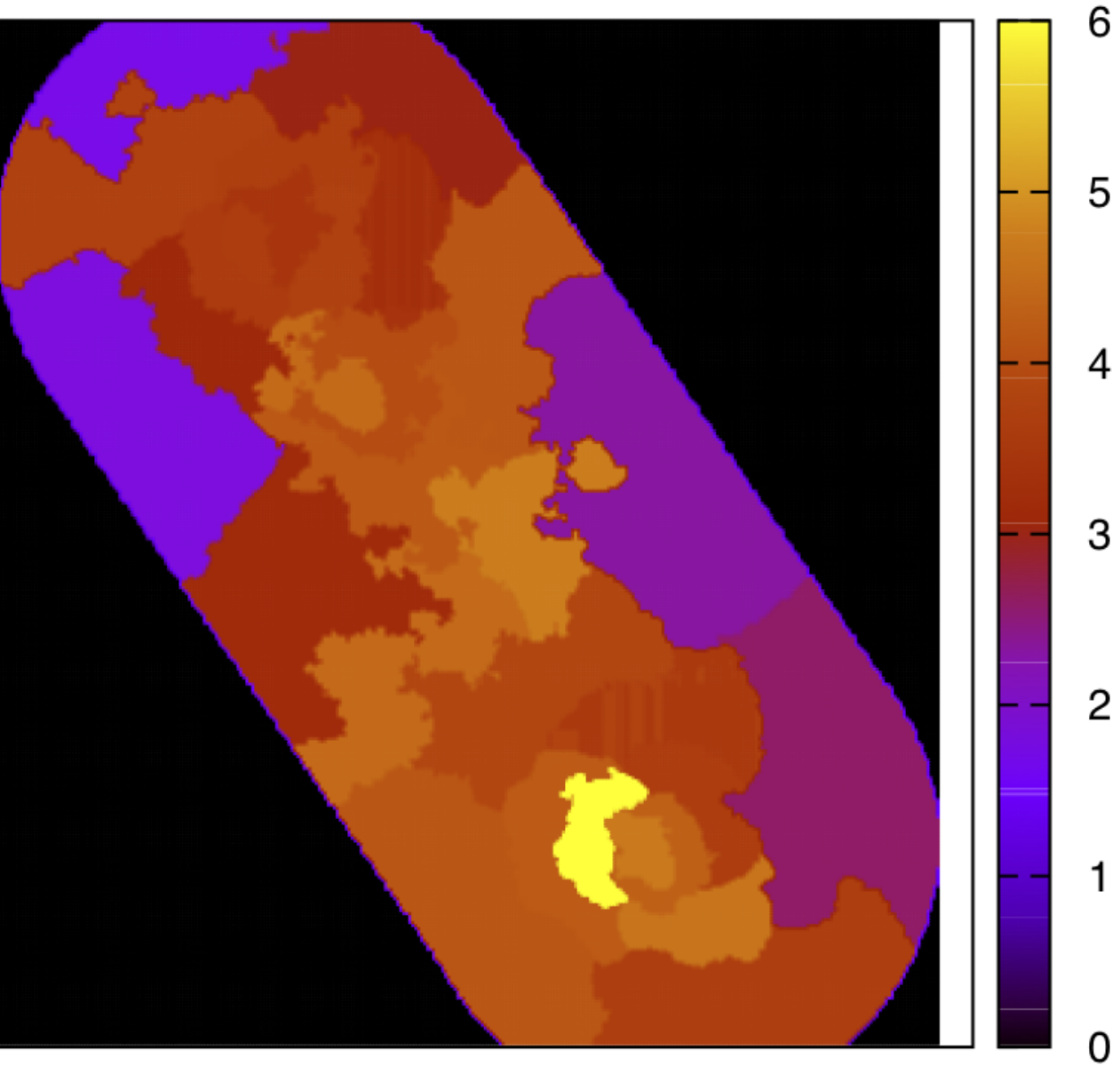}
          \includegraphics[width=.29\textwidth]{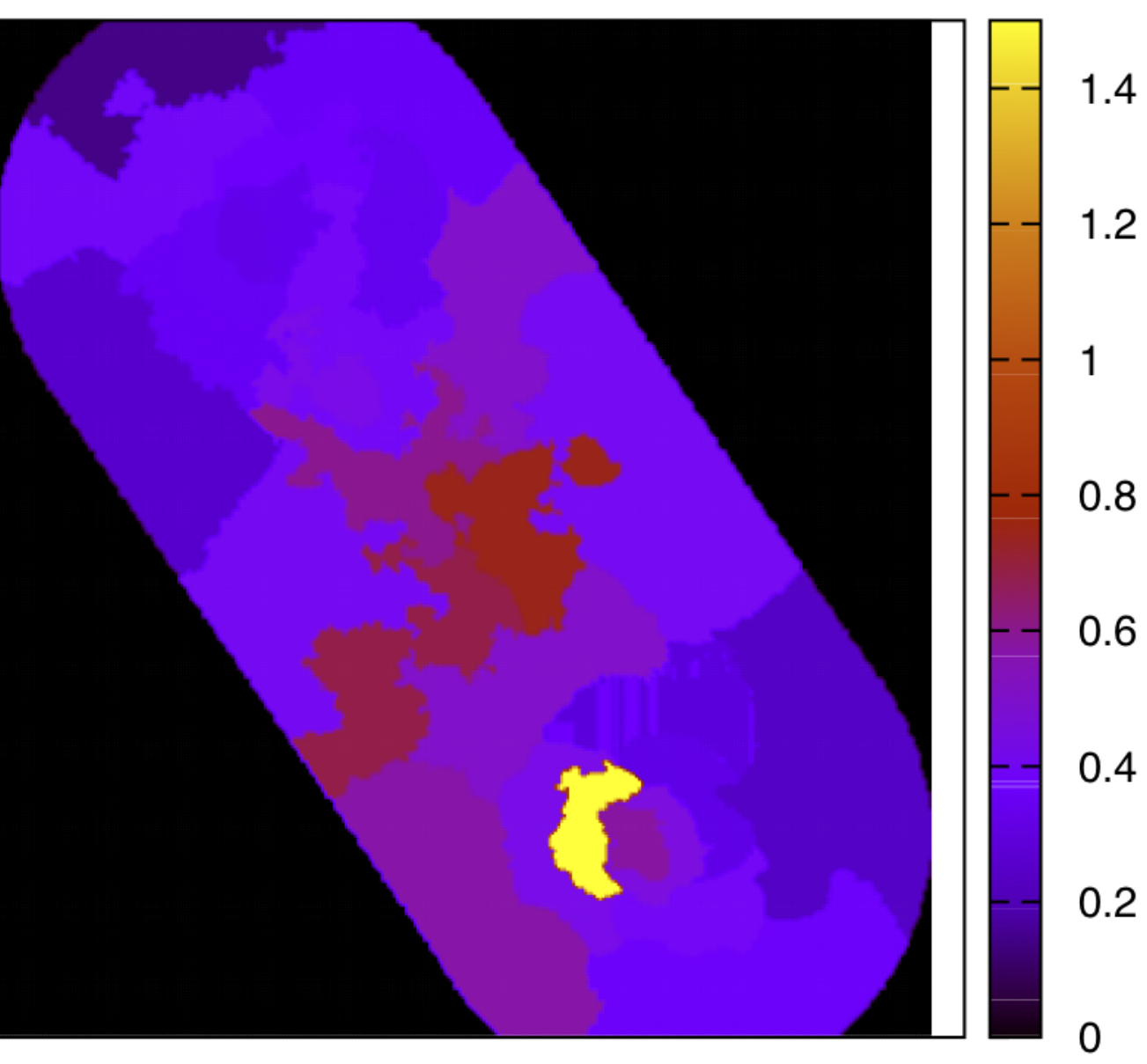}
          \includegraphics[width=.33\textwidth]{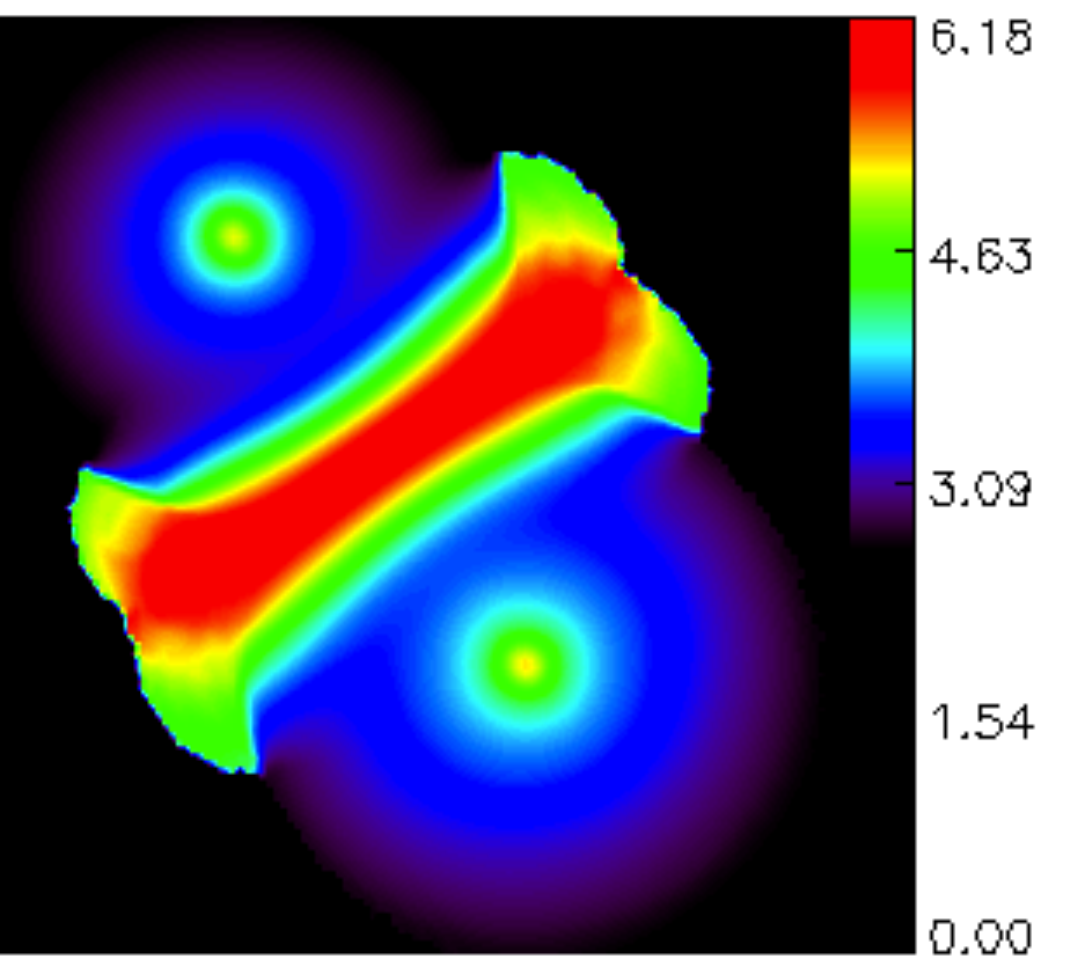}

   }
}
\caption{ \protect
 Images of X--ray surface brightness and projected temperature (in units of keV) 
 of A1750 from \CHANDRA\ observations and from our \FLASH\ simulation 
 with an impact velocity of 1400 \KMSEC\ (run: R3V14P15), 
 which is the best match to X--ray, optical and infrared observations.
 From left to right: 
 1st row:
 exposure corrected X--ray surface brightness map from observations,
 surface brightness map from simulation;
 2nd row: 
 projected temperature and error map from 
 observations and our simulation.
\medskip
\label{F:XRAY_COMPARE}
}
\end{figure*} 

In Figure~\ref{F:TSPEC_OBS} we show our results on the projected X--ray
temperature as a function of distance defined along the line
connecting the two X--ray peaks (the centers of the two main
components), setting the origin at the center of A1750N.  
The data points (squares) with error bars represent the results from a 
simultaneous fit to both \CHANDRA\ observations (ObsIDs 11878 and 11879).
Fits to individual observations are consistent with the simultaneous
fit, although ObsID 11879 show larger dispersion. 
Our results from the simultaneous fitting are in good
agreement with those based on \XMM\ observations
\citep{Belset2004AA415}.

We find that the projected model temperature keeps increasing towards
the center of A1750C (in agreement with Belsole et al.) implying that
is not a cool core cluster, since the projected temperature keeps
increasing towards the center, unlike A1750N, which has a decreasing
temperature profile towards its center (see Figure~\ref{F:TSPEC_OBS}).
The temperatures at about 0.2 Mpc from the centers of the two
components located at 0.2 and 0.7 Mpc, are about 3.5 and 4 keV respectively,
increasing toward the middle of the interaction region (0.3--0.6 Mpc)
to about 6 keV.  This is a clear sign of a compressed, shock heated
intracluster gas between the two merging clusters. This high
temperature region is clearly not in hydrostatic equilibrium in the
gravitational potential of A1750.

%
%
 \begin{table}
 \centering
 \caption{Initial Parameters and rotation angles}
 \begin{tabular}{|c|r|r|r|} \hline
    ID                            &  V[km s$^{-1}]$   &     P[kpc]     &  $\theta[^{\circ}]\;\;\;\;\;$     \\ \hline \hline 
    R1V05P15               &         500             &       150       &             16.9$\pm$2.6        \\ \hline
    R2V08P15               &         800             &       150       &             21.1$\pm$3.2        \\ \hline
    R3V14P15               &       1400             &       150       &             28.2$\pm$4.3       \\ \hline
    R4V16P15               &       1600             &       150       &             30.2$\pm$4.8         \\ \hline
    R5V35P15               &       3500             &       150       &             46.0$\pm$7.2         \\ \hline
    R6V08p00               &         800             &           0       &             21.2$\pm$3.4         \\ \hline
    R7V08p20               &         800             &       200       &             24.6$\pm$3.9         \\ \hline
 \end{tabular}
 \label{T:TABLE1} 
 \tablecomments{See text for explanation and the other initial parameters.
\medskip
}
\end{table}                                                          \medskip

\smallskip
\section{FLASH Simulations}
\label{S:FLASH}

We carried out three--dimensional self-consistent numerical
simulations of binary galaxy cluster mergers including dark matter and
intracluster gas using the publicly available parallel Eulerian
parallel code,
\FLASH, developed at the Center for Astrophysical Thermonuclear Flashes
at the University of Chicago (for a detailed description see
\citealt{Fryxell2000ApJS131p273} and \citealt{Ricker2008ApJS176}).  The
highest resolution (cell size) of 12.7 kpc we reached at the cluster
centers and in the regions of merger shocks.  Our box size was 13.3 Mpc on a
side. Our simulations were semi--adiabatic in the sense that only
adiabatic processes and shock heating were included (no other
non--gravitational effects).  A detailed description of the set up for
our simulations can be found in \cite{Molnet2012ApJ748}.  
We briefly summarize the main points here.

\subsection{Initial Conditions}
\label{SS:INIT}

Spherical symmetry is assumed for each of the two interacting clusters
for both the gas and the dark matter within the the virial radius of each cluster.  
We used a truncated NFW model \citep{NFW1997ApJ490p493} for the dark matter distribution:
\begin{equation} \label{E:NFW}
      \rho_{DM} (r) =  { \rho_s  \over x (1 + x)^2}
,
\end{equation}
\nop
where $\rho_s$ is the scaling parameter for the density, $x = r/r_s$, where
$r_s = R_{vir}/c_{vir}$ is the scaling parameter for the radius, and 
$c_{vir}$ is the concentration parameter.
We assume a truncated non-isothermal $\beta$ model for the gas:

\begin{equation}  \label{E:NFW}
      \rho (r) =  { \rho_0  \over (1 + y^2)^{3 \beta /2} }
,
\end{equation}
\nop 
where $y = r/r_{core}$, and $\rho_0$, $r_{core}$ are the central
density and gas scale radius, and $r \le R_{\rm vir}$.  The
temperature of the gas was determined form the equation of \HE\ via
numerical integration to ensure an initial equilibrium configuration
for each of the two main mass components.  The equation of state for the
gas was an ideal gas equation of state with $\gamma = 5/3$.  We
adopted $r_{core} = 0.08$ \RVIR, and, since we are interested in the
outer parts of the intracluster gas, $\beta=1$ (which is suggested by
cosmological numerical simulations,
\citealt{Molnet10ApJ723p1272}).

We adjust the initial concentrations of the dark matter components in order to
approximately reproduce the gas temperature profile lying well away
from the gas interaction zone, as described in the next
subsection. Our dark matter particles also represent stellar matter in
galaxies as this component may also be considered collisionless.  The
number of dark matter particles at each AMR cell was determined by the
local density, with the total number of 5 million dark matter particles used in
our simulations. In our simulations, we assume a gas mass fraction of 0.14
initially.

The velocities of the dark matter particles were determined by
sampling a Maxwellian distribution with the velocity dispersion,
$\sigma_r$, derived from the Jeans equation
\citep{LokasMamon2001MNRAS321}. Assuming isotropic velocity dispersion 
(the angular and radial components are equal: $\sigma_\theta = \sigma_r$), 
and NFW models, we obtain 

\begin{equation}  \label{E:SIGMA}
    \sigma_r^2 (r) = V_v^2 \, g(c_{vir}) \, x \, q^2 \int_x^\infty  
                           \Biggl[ {  \ln q \over  x^3 q^2 } - {  c_{vir} \over x^2 q^3 } \Biggr]  dx
,
\end{equation}
where $q = 1 + x \, c_{vir}$, the circular velocity is $V_v^2 = G M_{vir} / R_{vir}$, and 

\begin{equation}  \label{E:GC}
    g(c_{vir})  =  \bigl[ \ln (1 +  c_{vir} )  -  c_{vir} / ( 1 +  c_{vir} ) \bigr]^{-1}
.
\end{equation}
The direction of the velocities were assumed to be isotropic (for a
more detailed description see \citealt{Molnet2012ApJ748}).

%
%
\begin{figure}
\centerline{
\includegraphics[width=.48\textwidth]{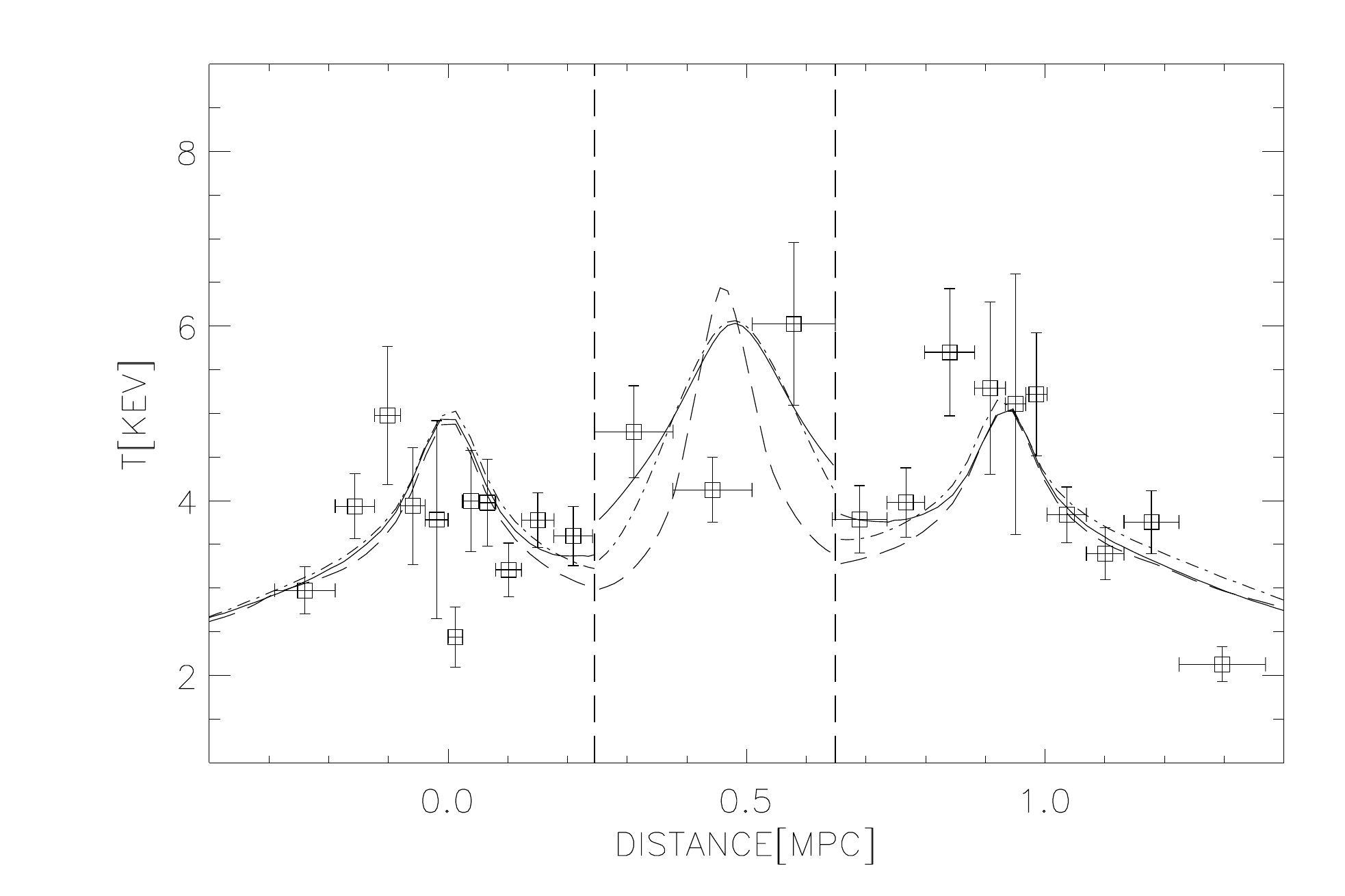}
}
\caption{ \protect
Projected X--ray temperature in A1750 from \CHANDRA\ observations (points
with error bars) and from \FLASH\ simulations 
(runs: R1V05P15, R3V14P15 and R5V35P15) 
from radial half circle regions around the two
main clusters and rectangle regions between them (regions shown in
Figure~\ref{F:CHANDRA}).  We sample a range of impact velocities of
500, 1400, and 3500 \KMSEC\ (continuous, dash--dotted, and dashed lines). 
The region between the vertical dashed lines represent the extraction region 
for the spectra for the interaction region (region 3 in Figure~\ref{F:CHANDRA}).
\label{F:TSPEC_VEL}
}
\end{figure} 

\subsection{FLASH Runs of Merging Clusters}
\label{SS:RUNS}

We performed simulations to find the best model for A1750, and, in general, to identify 
initial parameters we can determine for a merging cluster before the first core passage. 
Our main focus is on determining the impact velocity of the merger, because of its
relevance for cosmology.

We use gravitational lensing measurements as our starting point.
These lensing measurements are very important when using numerical 
simulations to interpret observations of merging galaxy clusters because 
they can be used to identify and constrain the main mass components of the system. 
In particular, the morphology of the reconstructed mass surface distribution 
tells us about how many components are in the system, and the projected positions
of the components give us information about the phase of the collision
(i.e., the time before or after first core passage).  Perhaps even more
importantly, we can get mass estimates for the components, which can
reduce the initial parameter space substantially we need to search
when we run a set of simulations.

Firstly we run a series of simulations of galaxy cluster mergers with
different concentration parameters and with the masses of each of the
two main components covering 1--5 \TMSUNFOUR. 
This is chosen to be roughly within the range allowed by the lensing analysis
of \cite{Okab2008PASJ60}, except that we did not consider masses larger
than 5 \TMSUNFOUR\ because their virial radii would be too large and therefore
the temperature increase in the interaction region would be much greater than 
what observed 
(no useful constraints on concentration parameters are available from lensing). 
We find that assuming total masses, 
$M\,^C = 2.0$ and $M\,^N = 1.8$ \TMSUNFOUR\ 
and concentration parameters of 8 and 10 for the the main and the
infalling cluster, we can reproduce the projected temperature profiles
of the central regions of both clusters in the undisturbed cluster
regions, away from the gas interaction zone in between the clusters.

%
%
\begin{figure}
\centerline{
\includegraphics[width=.48\textwidth]{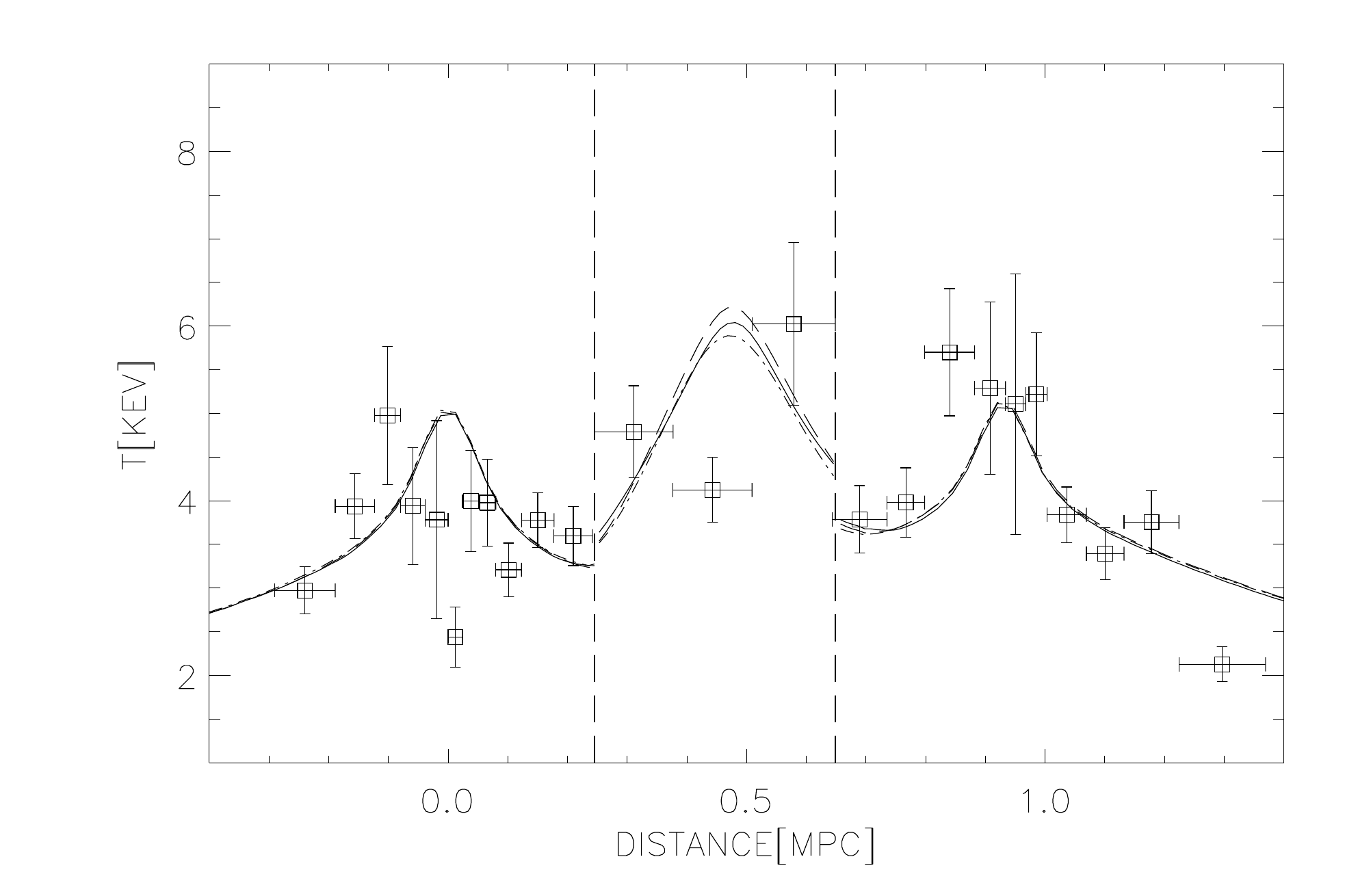}
}
\caption{ \protect
Same as Figure~\ref{F:TSPEC_VEL}, but from 
our \FLASH\ simulations with different impact parameters of 0, 150, and 200 kpc 
(dashed, continuous, and dash--dotted lines; runs: R6V08p00, R2V08P15, and 
R7V08p20).
\label{F:TSPEC_PIMP}
\medskip
}
\end{figure} 

In the next step, we hold the two masses and concentration parameters
fixed at these values, and run a set of simulations systematically
changing the impact parameters and relative velocities. In
Table~\ref{T:TABLE1} we summarize the initial parameters for those
runs we discuss in detail in our paper.  In this table the first
column is the identification number for our runs using the following convention: 
the numbers after V and P represent the initial relative velocity in units of 
100 \KMSEC, and the impact parameter in units of 10 kpc.
In the second and third columns we list the initial velocities and
impact parameters, the third column shows the angle, $\theta$, we
rotated the system out of the plane of the sky towards the observer
(along an axis which is perpendicular to the line connecting the two
cluster centers) in order to reproduce the observed projected distance
between the two X--ray peaks.

%
%
\begin{figure*}
\centering{
   \subfigure{
          \includegraphics[width=.33\textwidth]{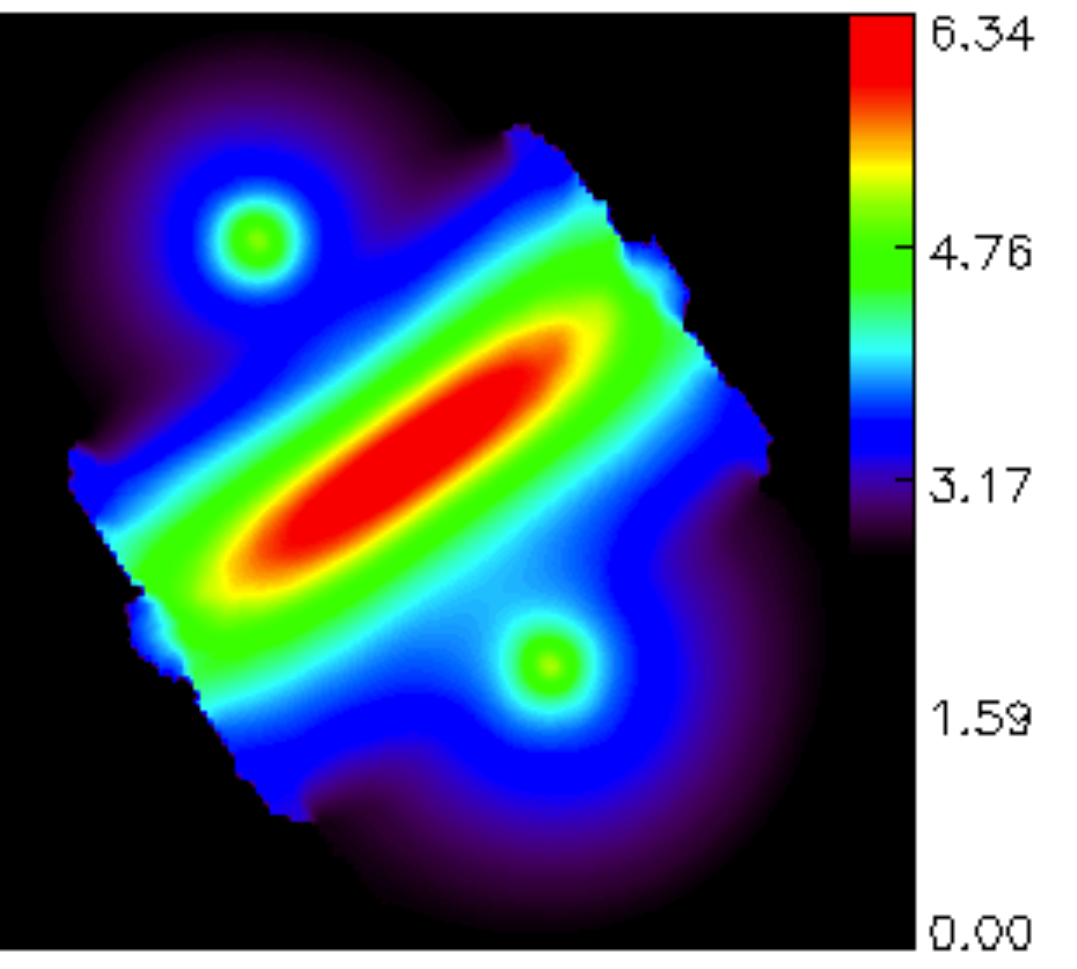}
          \includegraphics[width=.33\textwidth]{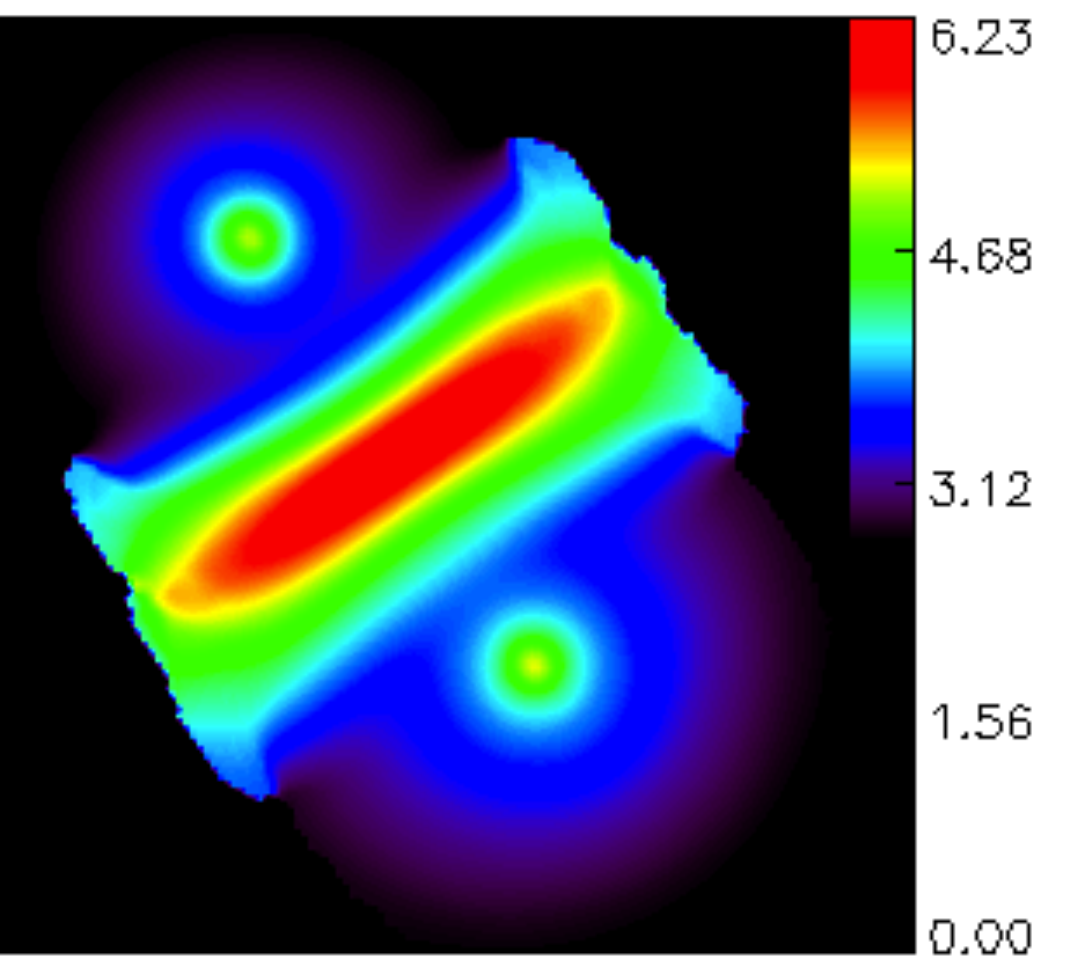}
   }
   \subfigure{
          \includegraphics[width=.33\textwidth]{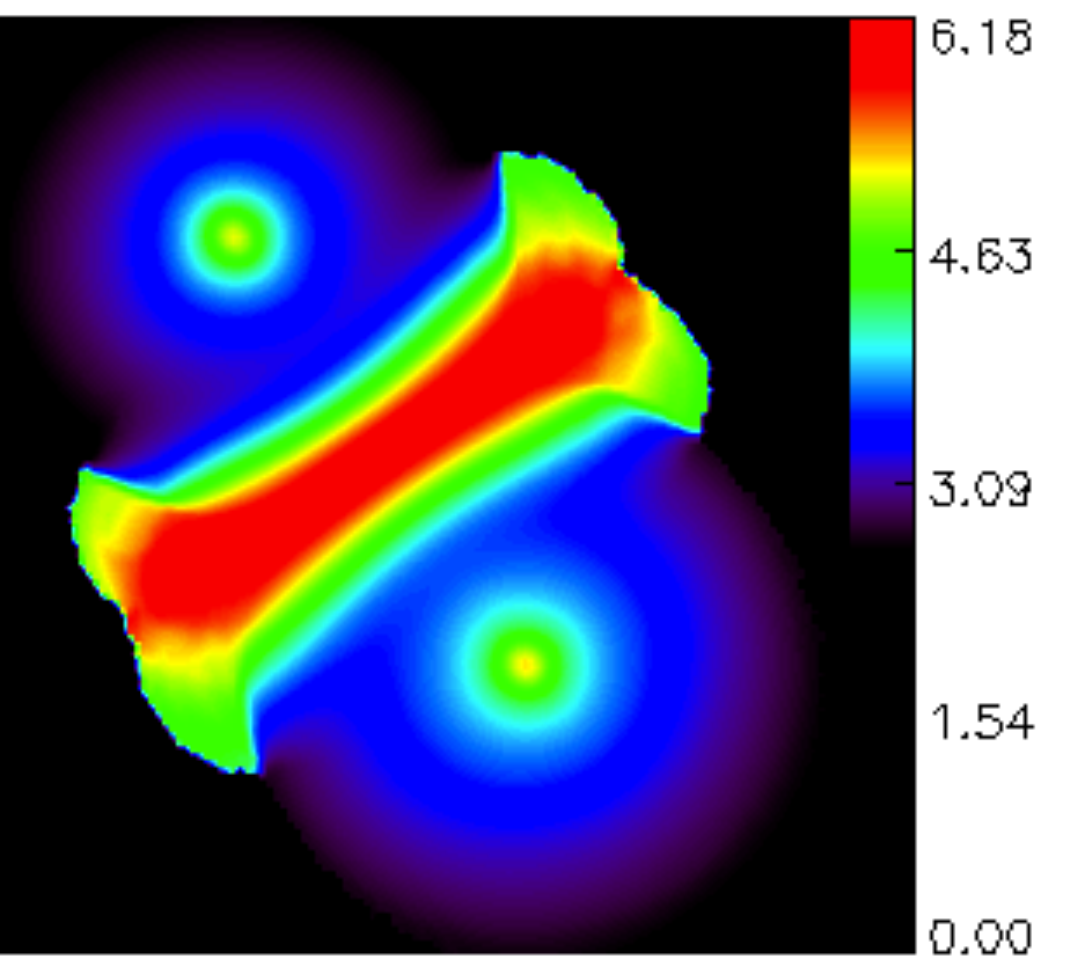}
          \includegraphics[width=.33\textwidth]{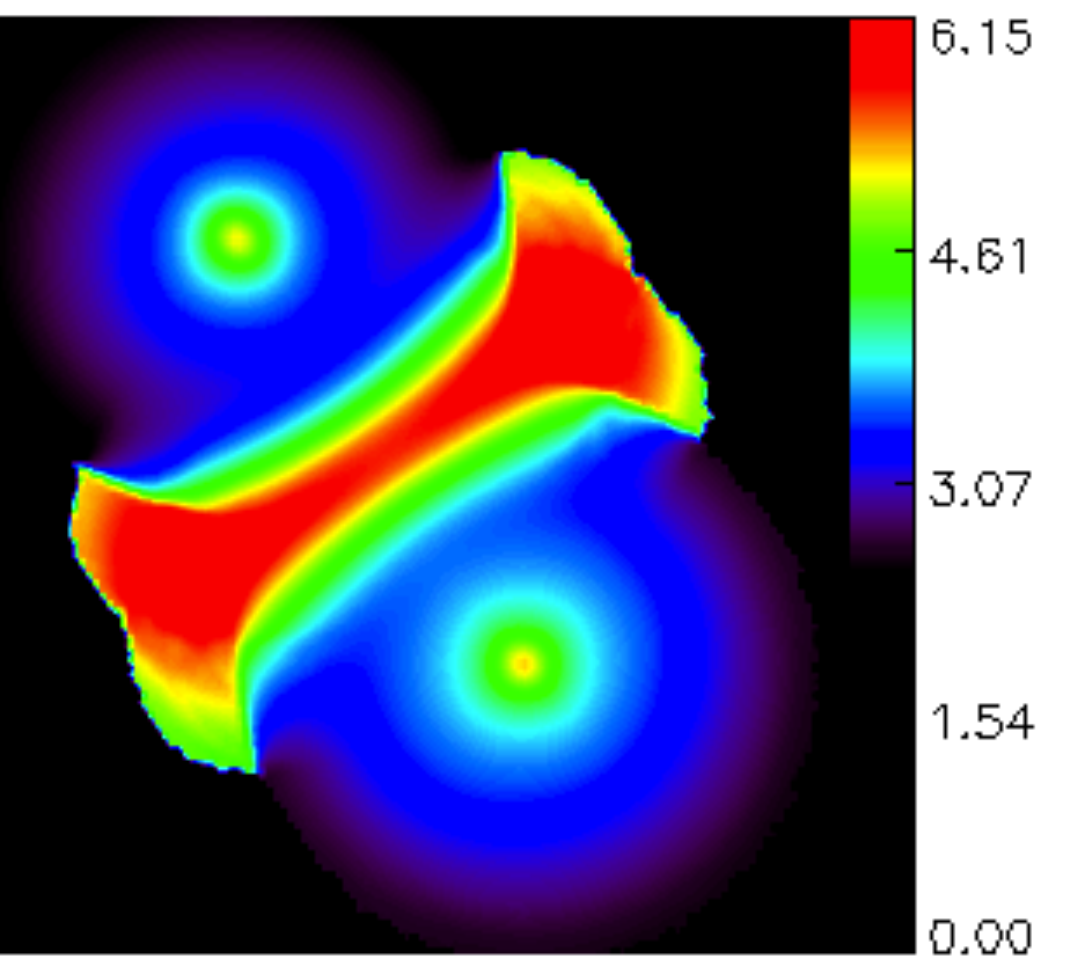}
          \includegraphics[width=.33\textwidth]{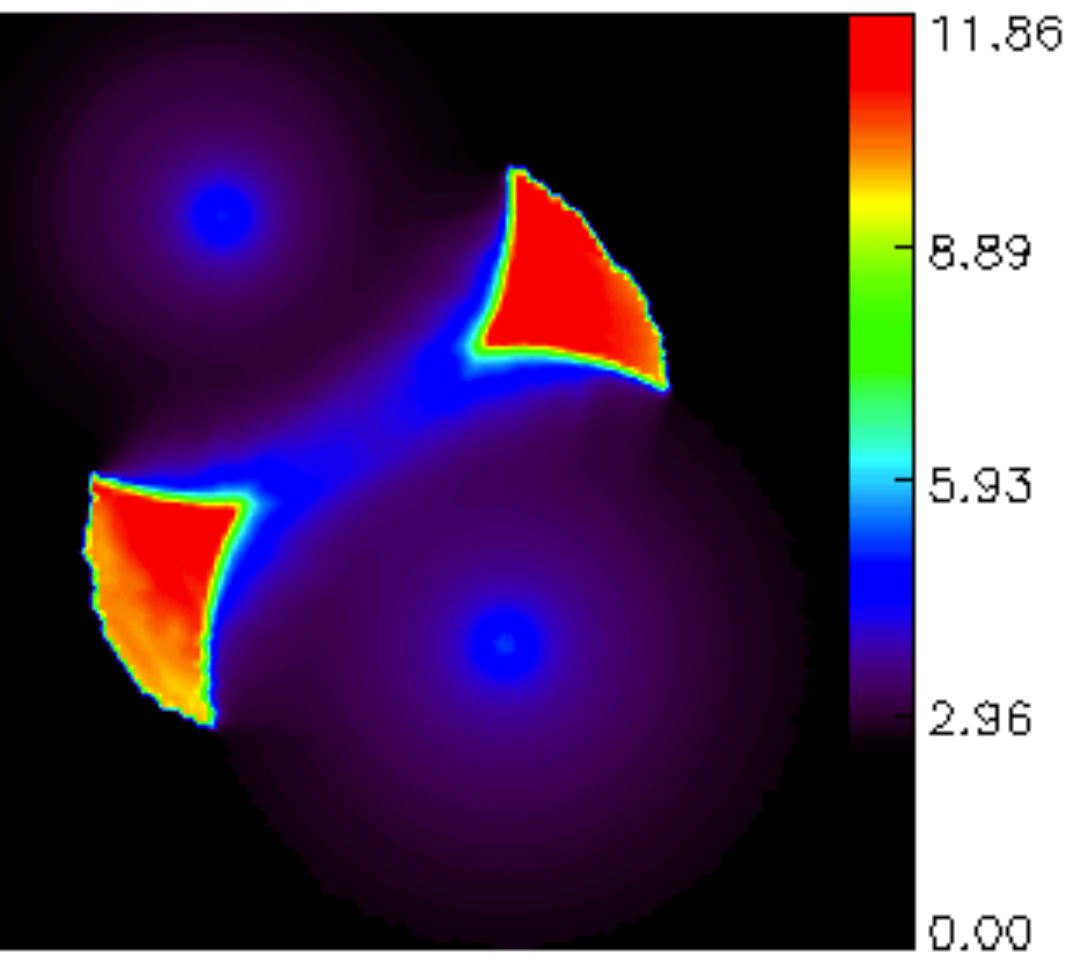}
   }
}
\caption{ \protect
 Projected X--ray temperature maps from our \FLASH\ simulations in units of keV, 
 which provide a good match to the observed projected temperature distribution in A1750
 (see Figure~\ref{F:TSPEC_VEL}).
 From left to right: temperature maps from simulations with 
 initial velocities of 500 and 800 \KMSEC\ (runs: R1V05P15 and R2V08P15, 1st row),
 and with 1400, 1600, and 3500 \KMSEC\ 
 (runs:  R3V14P15, R4V16P15, and R5V35P15, 2nd row).
 The interaction regions with an enhanced temperature of about 6 keV (red color) 
 can be clearly seen in these images between the peaks corresponding to the two subclusters 
 (except in the last one, where the maximum temperature is 10--11 keV).
\medskip
\label{F:TSPEC_SIM}
}
\end{figure*} 

\smallskip
\subsection{Simulated Images}
\label{SS:IMAGES}

After each simulated collision is completed, we generate X--ray
surface brightness and projected X--ray temperature images for a range
of viewing angles.  We align the two mass centers with the $x$
coordinate axis choosing the $y$ axis so that the $x-y$ plane
coincides with the plane of the collision, which we first consider to
be the plane of the sky. Then, we rotate our cluster around the $x$ axis
with a roll angle, $\phi$, (active rotation), then around the $y$ axis
by a polar angle of $\theta$. 
Therefore, with this definition, zero rotation angles ($\phi = 0$, $\theta = 0$) 
mean that the collision is in the plane of the sky.
Once we choose the roll angle and the polar angle, we generate the
X--ray surface brightness image by performing a line--of--sight (LOS)
integral
\begin{equation}  \label{E:XRAY}
    I_X (x,y) \propto  \int_{\ell_1}^{\ell_2}
                           \rho_g^2 \, \Lambda (Z_{ab}, T_g) \; d \ell
,
\end{equation}
where $\ell$ is the spatial coordinate in the LOS, and 
we used the publicly available \APEC\ code
\footnote{http://www.atomdb.org}
for the frequency integrated cooling function, $\Lambda$, since in the early 
phase of the merging (well before first core passage), the temperature does 
not get above about 12 keV, therefore we do not need relativistic corrections.
We convolve the derived total spectrum in each pixel of the resulting
map with a piecewise linear approximation of the on--axis affective
area of the 
ACIS--I CCD\footnote{http://asc.harvard.edu/proposer/POG/html/ACIS.html},
to derive the final simulated image of the X--ray surface brightness.

We approximate the observed projected X--ray temperature for our
merging clusters using the spectroscopic-like X-ray temperature
proposed by \cite{Mazzet04MNRAS354}, which is a weighted average of
the physical temperature in the LOS,
\begin{equation} \label{E:TSPEC}
       T(x,y) \approx T_{sp} (x,y) =  { \int w_{sp} \; T_g \; d \ell \over \int w_{sp}\; d \ell}
,
\end{equation}
where the weight is $w_{sp} = \rho^2 / T_g^{3/4}$.
This projected temperature has been shown to be a good approximation to cluster 
temperature profiles similar to those observed with 
\CHANDRA\ and \XMM\ \citep{Nagaet07}.

\smallskip

\smallskip
\section{Results and Discussion}
\label{S:Results}

Our results for the simulated projected temperature distributions for
different impact velocities as well as different impact parameters are
shown in Figures~\ref{F:TSPEC_VEL} and \ref{F:TSPEC_PIMP}.  In
Figure~\ref{F:TSPEC_VEL} we show projected X--ray temperatures from
simulations with impact velocities of 500, 1400, and 3500 \KMSEC\
(continuous, dash--dotted, and dashed lines). 
The region between the two vertical dashed lines represents
the interaction region.  The rotation angle out of the plane of the
sky, $\theta$, and the phase of the collision (the time before first
core passage) are set such that the projected distance is equal to the
observed separation of 900 kpc, and the projected temperature profile
simultaneously matches the profile we derived from X--ray observations (see
Table~\ref{T:TABLE1}).  Note that we did not aim to model the cool core of
A1750N since we are interested in the interaction region, thus we do
not expect to reproduce the projected temperature in the core.

We focus on modeling the interaction region, which is between the two
temperature peaks excluding the core regions around the clusters
within a radius of about 0.2 kpc, which is, in our distance
coordinate, from 0.15 to 0.8 Mpc.  We obtain acceptable fits in the
interaction region to all impact velocities we considered, except the
high impact velocity of 3500 \KMSEC, for which the simulated
temperature profile deviates more than 2$\sigma$ from all three data
points in this region (three points in the middle).  Based on this
figure, we conclude that a large impact velocity for A1750 is very
unlikely.

We show the effect of different impact parameters in
Figure~\ref{F:TSPEC_PIMP}.  It can be seen from this figure that
varying the impact parameter over 0, 150, 200 kpc does not change the
projected temperature significantly. Clearly, the projected
temperature is not sensitive to a moderate change in the impact
parameter, therefore we can not put meaningful constraint on this
parameter.  Therefore we use our results for runs with fixed impact
parameter at 150 kpc in the following discussion.

As a next step, we check if we can use the morphology of the projected
X--ray temperature to further constrain the impact velocity of A1750.
We show images of the exposure corrected X--ray surface brightness and 
projected temperature based on the merged \CHANDRA\ observations of A1750
with ObsIDs 11878 and 1187 in Figure~\ref{F:XRAY_COMPARE}. 
The enhancement between the two main clusters can be clearly seen
in the surface brightness image (1st row 1st panel). 
In the 2nd row 1st and 2nd panels we show the images of the projected temperature 
and the corresponding error using adaptive binning of the image based on surface brightness 
contours following the method described by \cite{Sanders2006MNRAS371}.
The errors in the temperature in the interaction region are large, about 1 keV, 
due to the low surface brightness and the short exposure time.
The surface brightness contours in this region are dominated by noise, 
therefore the temperature map is not reliable (the projected temperature map
in the interaction region based on the merged observations do not agree with 
those from on the individual observations). 
We can only conclude that there seem to be no large temperature peaks or troughs 
in this region. Our surface brightness and projected temperature 
temperature maps are consistent with those of \cite{Belset2004AA415},
who used a 34 ks \XMM\ observation of A1750 (their Figure~2 and 5). 
The projected temperature map of A1750 from \XMM\ observations has
an error of about 0.6 keV in the interaction region showing a temperature
enhancement with no significant increase or decrease towards the middle of 
the region.

The images of the projected temperature from our \FLASH\ simulations with impact
velocities of 500, 800, 1400, 1600, and 3500 \KMSEC\ 
(runs R1V05P15, R2V08P15, R3V14P15, R4V16P15, and R5V35P15; 
left to right, 1st and 2nd row) are shown in Figure~\ref{F:TSPEC_SIM}.
The rotation angles and phases of collision are different 
for these runs chosen to match the observed projected distance and
temperature profile of A1750 (see Table~\ref{T:TABLE1} and Figure~\ref{F:TSPEC_VEL}). 
The two peaks on the lower right and upper left corner in each image 
mark the cluster centers.  
The enhanced projected temperature region in the middle is the interaction 
region. Since we need larger rotation angles out of
the plane of the sky for larger velocities in order to have more
cooler gas in the LOS to lower the projected temperature to match with
the observations, the different morphology is due to the different
rotation angles. Therefore, as we can see it from Figure~\ref{F:TSPEC_SIM},
in general, the temperature distribution in the interaction
region has a bump in the middle for low impact velocities 
and turns into a saddle 
shape for larger velocities as a function of the rotation angle, $\theta$.

Our simulations assuming an impact velocity of 1400 \KMSEC\ 
and 1600 \KMSEC\ (runs R3V14P15 and R3V16P15) with a flat plateau of 
about 6 keV in the projected temperature in the interaction region 
(1st and 2nd panel in the 2nd row in Figure~\ref{F:TSPEC_SIM}) 
have the most similar morphology to those from our \CHANDRA\
analysis (2nd row 1st panel in Figure~\ref{F:XRAY_COMPARE}) and from 
the \XMM\ observation of A1750 (Figure~5 of \citealt{Belset2004AA415}).  
There are some small deviations due to substructure we have not modeled 
in our simulations, but the interaction region looks very similar. 
In this region R3V14P15 and R3V16P15 show different morphology 
at the level of about 0.5 keV, but these systematic variations are not significant,
they are in the same order as the errors from observations.
Therefore, unfortunately, we can not derive an accurate impact velocity
for A1750 based on the available projected temperature maps, 
because neither the \CHANDRA\ nor the \XMM\ observations have sufficient 
exposure time to construct an accurate projected temperature map.
However, in principle, the morphology of the projected temperature can be used 
to put strict constraints on the impact velocity for merging clusters.
Our results from simulations suggest that if we have observed 
projected temperature maps with errors less than 0.5 keV, we may be able to
constrain the impact velocities of merging clusters to 200--300 \KMSEC\
(compare 1st to 2nd panels in the 1st  as well as in the 2nd row in 
Figure~\ref{F:TSPEC_SIM}).

The Mach number for our best fit model (run R3V14P15, see
Figure~\ref{F:XRAY_COMPARE}) can be calculated from our simulations
directly.  We show the physical temperature and pressure profiles of
the gas along a line crossing the shocks as a function of distance
(same spatial coordinate as in Figure~\ref{F:TSPEC_VEL}) in units of
keV and, 4$\times 10^{11}$ dyne cm$^{-2}$, 
to be able to see their features in the same plot for easy comparison.  
The temperature peak of about 11 keV in the middle of the interaction region 
is due to previous shock activity followed by adiabatic compression at the
contact discontinuity (no jump in the pressure here).  The temperature
and pressure jumps at -170 kpc and 180 kpc belong to the two shock
fronts we can see before the first core passage.  We calculate the
Mach number from the usual Rankine--Hugoniot jump conditions using the
pressure jump:
\begin{equation}
 {\cal M} =  \Biggl[ 0.8 \, \frac{P_2}{P_1} + 0.2 \Biggr]^{1/2}
,
\end{equation}
where $P_1$ and $P_2$ are the upstream (pre-shock) and downstream (post-shock)
pressures at the position of the shock, and we assumed $\gamma = 5/3$, as in 
\cite{Molnet09ApJ696}, where the advantages of using the pressure is discussed.
Since the pressure jumps are similar in both shocks, 
we get similar Mach numbers, 1.4 and 1.2, for the shock front on the left and the right.

%
%
\begin{figure}
\centerline{
\includegraphics[width=.48\textwidth]{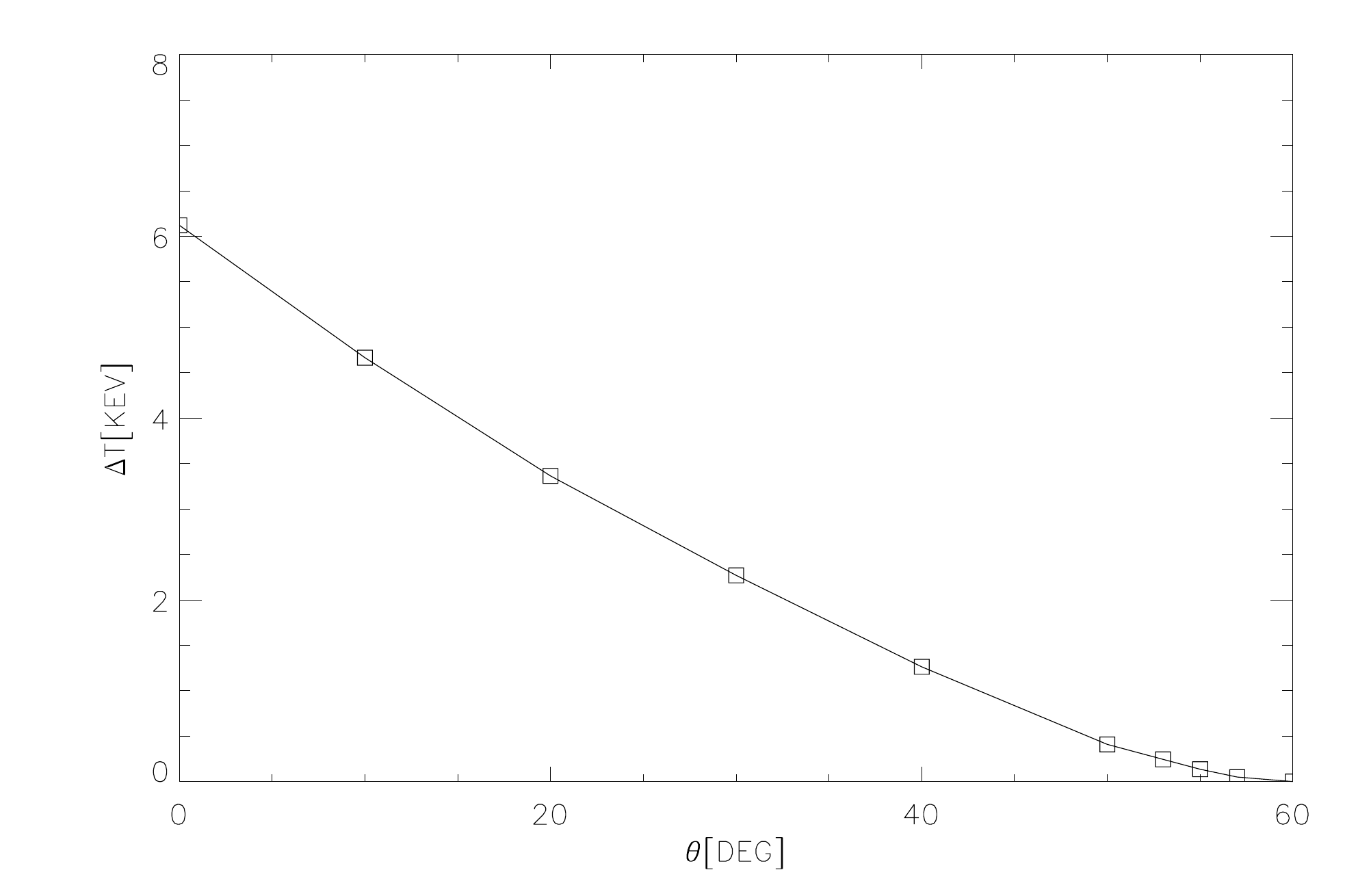}
}
\caption{ \protect
Projected temperature change as a function of rotation angle
($\theta$, in Degrees) at the center of the contact discontinuity for R3V14P15
(the same run and phase as in Figure~\ref{F:XRAY_COMPARE}).
In this Figure we show the temperature change 
along the line connecting the two cluster centers 
marked by the peaks in the projected
X--ray temperature maps (see Figure~\ref{F:TSPEC_SIM}).
\medskip
\label{F:SHOCKROTATE}
}
\end{figure} 

\begin{figure}
\centerline{
\includegraphics[width=.48\textwidth]{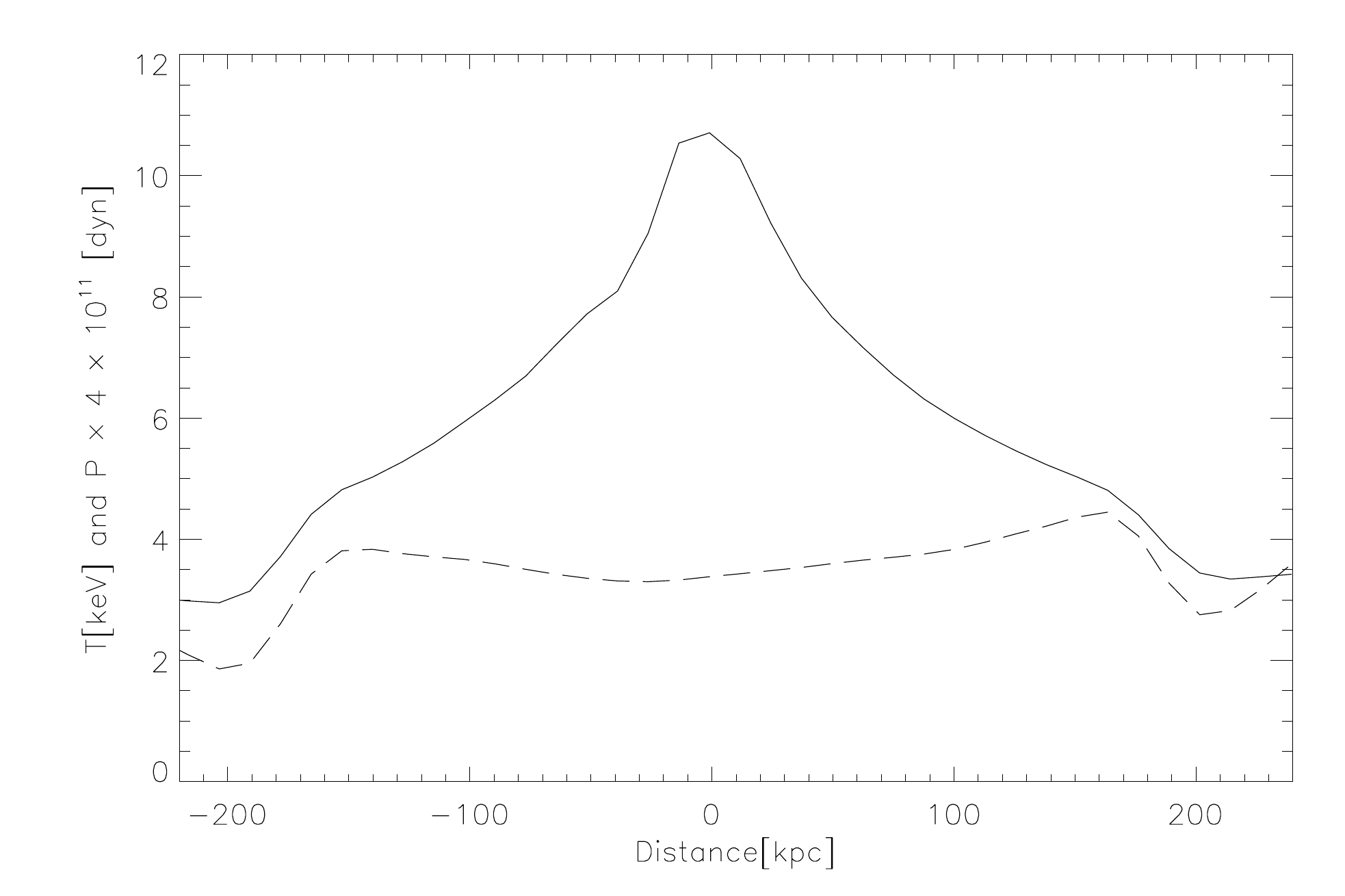}
}
\caption{ \protect
Physical temperature and pressure (solid and dashed line) across the
shock as a function of distance for run R3V14P15 (the same run and
phase as in Figure~\ref{F:XRAY_COMPARE}).  
The distance coordinate is the same as in Figure~\ref{F:TSPEC_VEL}).  
The shock can be clearly seen on the two sides of the interaction region: 
as a temperature and pressure jump. The maximum temperature is at the 
contact discontinuity, where there is no pressure jump.
\medskip
\label{F:SHOCK}
}
\end{figure}

\cite{Belset2004AA415} estimated the Mach number for the shock 
assuming that the gas is close to isothermal 
and in equipartition after the shock, suggested by \cite{Markevitch1999ApJ521}, 
in which case the Rankine--Hugoniot jump conditions imply that 
\begin{equation} \label{E:MACH}
    {\cal M} =   \sqrt { 3 {\cal C} \over 4 - {\cal C} }
,
\end{equation}
%
where the shock compression, ${\cal C}$, can be derived from
\begin{equation}  \label{E:MACHMAR}
  { 1 \over {\cal C} } = \Biggl[  4 \biggl( {T_2 \over T_1} - 1 \biggr)^2 + {T_2 \over T_1}  \Biggr]^{1/2} - 
                                   2 \biggl( {T_2 \over T_1} - 1 \biggr)
,
\end{equation}
assuming $\gamma = 5/3$. Belsole et al. obtained a Mach number of
1.64 assuming pre shock temperature of $T_1 = 3.1$ and for the post
shock temperature, the global temperature of the interaction region,
$T_2 = 5.1$.  As we can see this estimated Mach number is in a 
good agreement with that we derived from our simulations directly, but
this is only a coincidence since some of the temperature increase in
the interaction region comes from adiabatic compression and not from a shock. 
From Figure~\ref{F:SHOCK} we can see that a global 
temperature in the interaction region would be about 8 keV not 5.1 keV
as measured by \cite{Belset2004AA415} from \XMM\ data, and would 
result in an overestimate of the Mach number (2.4), but, as we pointed out,
the projected temperature is, in general, and in this region, an
underestimate of the physical temperature due to cool gas in the LOS.
We illustrate this point in Figure~\ref{F:SHOCKROTATE}).  In this
figure, we show how the projected peak temperature changes as a
function of rotation angle out of the sky ($\theta$) using the same
run (R3V14P15) and phase as we used in Figure~\ref{F:XRAY_COMPARE}. 
We find that the peak temperature on the line connecting the two cluster
centers (in this simulation) decreases substantially with angle from
about 6 keV at $\theta = 0$, to zero keV at about $\theta$ = 60\DEGREE.

%
%

\begin{figure}
\centerline{
\includegraphics[width=.48\textwidth]{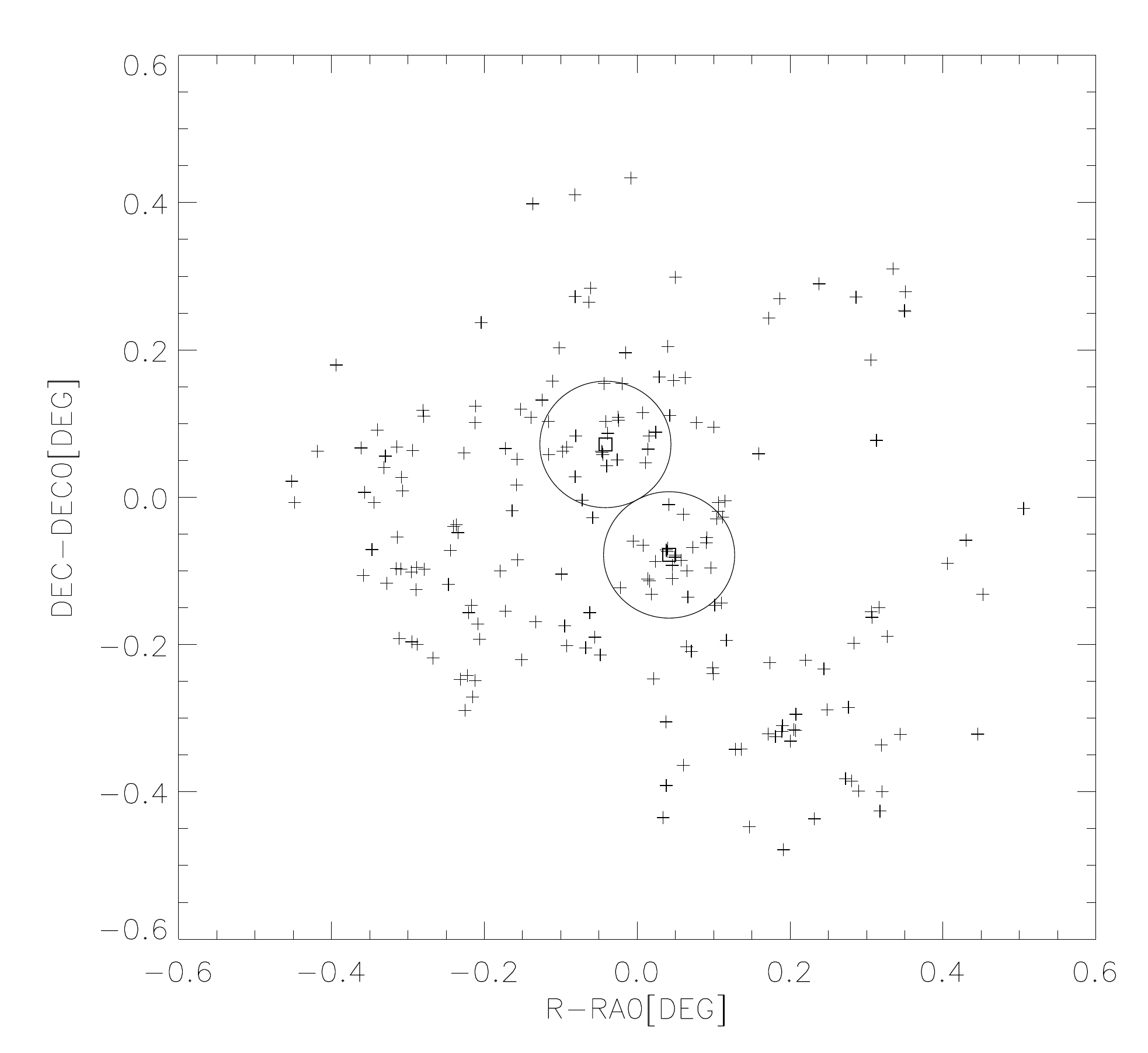}
}
\caption{ \protect
The distribution of galaxies in the field of A1750. The coordinates
are relative positions to the center of the image in units of Degrees
(plus signs). The centers of the A1750C (close to the center)
and A1750N (to the North) are marked with squares. 
The circles show the extraction regions we used in our histograms of 
velocity distributions (see Figure~\ref{F:REDSHIFT}).
The radii of the circles are equal to the half distance between the 
centers of the subclusters (about R$_{500}$). 
\label{F:GXSKY}
}
\end{figure}

\begin{figure}[t]
\centerline{
\includegraphics[width=.48\textwidth]{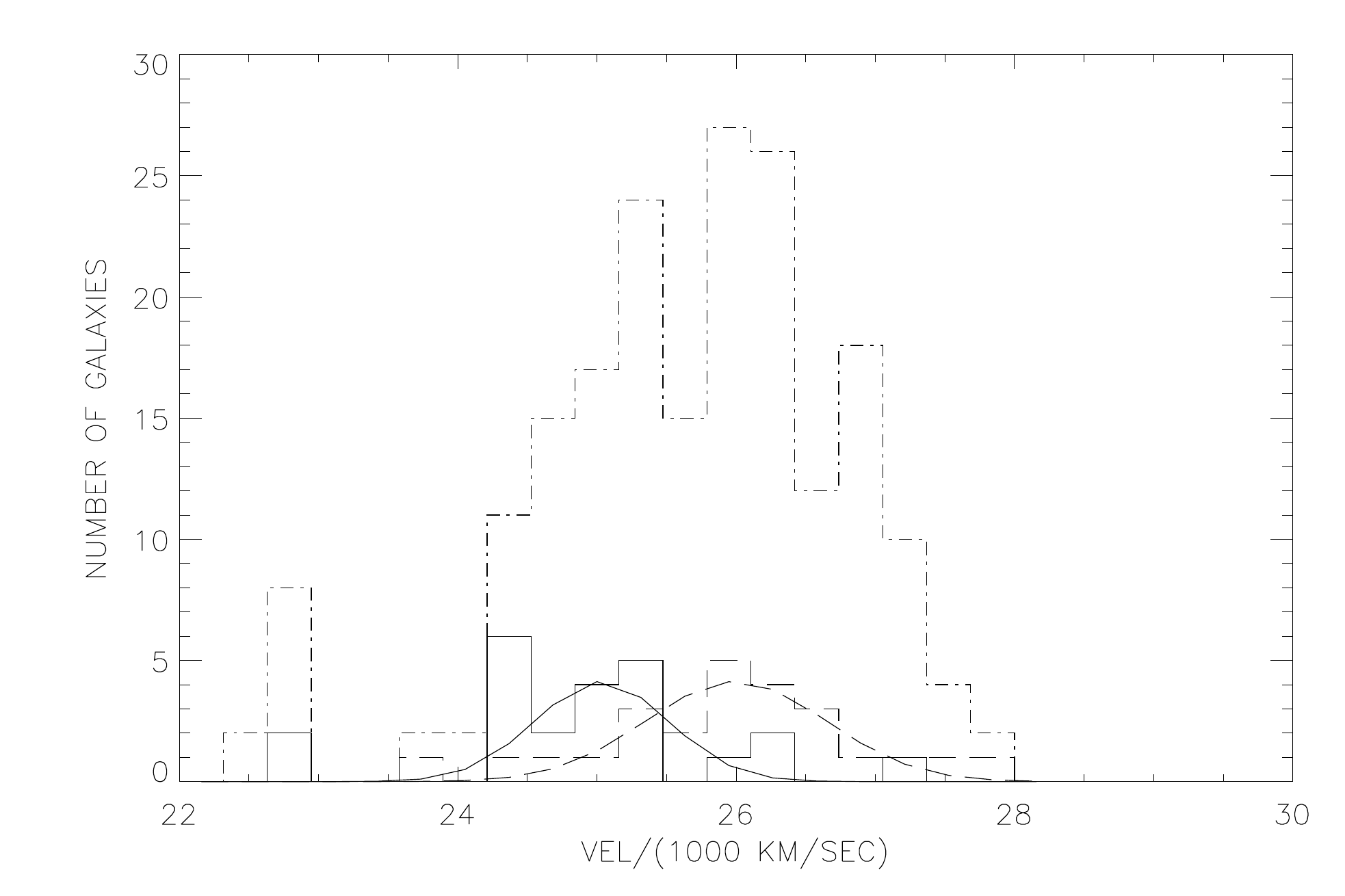}
}
\caption{ \protect
Velocity distribution of galaxies in the vicinity of A1750. 
The dash--dotted histogram represents all galaxies in the field (see Figure~\ref{F:GXSKY}).
Histograms with solid and dashed lines show distributions of velocities of galaxies within
circles around A1750N and A1750C (see Figure~\ref{F:GXSKY}). Gaussian fits to 
the histograms of the two subclusters are shown with the same line codes.
\label{F:REDSHIFT}
}
\end{figure}

As a last step, we investigate wether we can use galaxy redshift measurements 
in the field of A1750 to quantify the impact velocity in this merging cluster.
We collected galaxy redshift measurements in the vicinity of A1750 from
NED\footnote{http://ned.ipac.caltech.edu}.  We found 195 galaxies
within 30\SMALLER\AMIN\ from the center of A1750 with redshifts between
23000 \KMSEC\ and 28000 \KMSEC. We show the positions of galaxies in
the field of A1750 in Figure~\ref{F:GXSKY}. Squares mark the centers
of the two subclusters.  We show the number distribution of the
velocities of these galaxies in Figure~\ref{F:REDSHIFT}.  The
histogram with a dash--dotted line represents all galaxies, histograms with
dashed and continuous lines show distributions of galaxy velocities within a
radius of half of the distance between centers of A1750C and A1750N (about 1R$_{500}$).
We find radial velocities of 25999$\pm80$ \KMSEC\ and 25034$\pm101$ \KMSEC
for A1750C and A1750N respectively.
We can conclude that A1750C is closer to us moving towards A1750N, 
which is farther from us.  This is why the A1750C, which is closer to us, 
has a larger redshift.  As we can see, redshift measurements can break 
the degeneracy of the spatial distribution in the LOS inherent in our
X--ray and gravitational lensing observations.
To derive the projected instantaneous relative velocity from these redshift
measurements we fitted gaussians to the peaks of the observed velocity
distributions of the two components to derive their relative LOS
velocities. The results are shown in Figure~\ref{F:REDSHIFT} 
(solid and dashed Gaussians). The positions of the two peaks of the
Gaussian fits agree with the two local maxima of the histogram representing
the total galaxy distribution, as we would expect.
We obtained a LOS velocity of $V_r = 960\pm$130 \KMSEC. 
This should be considered to be a lower limit for the impact velocity for A1750 
because of the significant projection angle implied by the X--ray analysis above,
and because the relative velocity increases until the first core passage.

Using the fewer galaxy redshifts available at that time and adopting
a biweight analysis applicable to low-number statistics,
as suggested by \cite{BeersET1990AJ100}, \cite{HwangLee2009MNRAS397}
found radial velocities of 
25931$_{-201}^{+212}$ \KMSEC\ and 24999$_{-172}^{+260}$ \KMSEC\
for A1750C and A1750N, and thus obtained $V_r = 932\pm332$ \KMSEC\
(Table 2 in Hwang \& Lee).
The differences in the radial velocities of the two components and their relative velocities
between our results and those of Hwang \& Lee are only 68, 35 and 28 \KMSEC, 
each is only a small fraction of the 1$\sigma$ errors, 
thus we conclude that our results are in good agreement with those of Hwang \& Lee.

Comparing the instantaneous radial velocities of our simulations determined from the 
rotation angles derived from the projected temperature profiles 
(last column in Table~\ref{T:TABLE1}), 
we derive the initial relative velocity of 1460$_{-200}^{+180}$ \KMSEC\
(uncertainties in the rotation angle and radial velocity contributing 
about 130--150 \KMSEC\ each).
In Figure~\ref{F:XRAY_COMPARE} (1st row 2nd panel and 2nd row 3rd panel), 
we show the images of the X--ray surface brightness and projected temperature 
from our \FLASH\ simulation with an impact velocity of 1400 \KMSEC\ 
(run: R3V14P15)
which is the best match to X--ray, optical and infrared observations of A1750.
The enhancement in the surface brightness within the two X--ray peaks marking 
the centers of the merging clusters can be clearly seen in the simulated image 
(1st row 2nd panel), and similarly to the \CHANDRA\ surface brightness image 
(1st row 1st panel).
In the image of the projected X--ray temperature from our simulations 
(2nd row 3rd panel), the maximum temperature (red) in the middle of 
the interaction region is about 6 keV, which is consistent with our temperature map 
from the \CHANDRA\ observations (2nd row 1st panel) within errors.
Based on this figure, we conclude that the morphology of the simulated 
images of the X--ray emission and the projected temperature are in good 
agreement with those from our \CHANDRA\ analysis (1st row 1st panel and 
2nd row 1st panel) and those based on \XMM\ observations of A1750 
(images in Figure 5 of \citealt{Belset2004AA415}).

%
%

We compare our results for the velocity and 
rotation angle from simulations to those obtained applying the two-body dynamical model 
introduced by \cite{BeersET1982ApJ257}.
This simplified model for binary galaxy cluster mergers assumes that the two components
start out at time zero with zero spatial separation on a radial orbit ignoring a likely
finite angular momentum of the system and any tidal forces due to the large scale structure.
Since the X--ray observations suggest that the two subclusters are moving towards
each other, we are looking for a bound solution.
In this case the system can be described by the well known parametric solution of 
Einstein's field equations:
\begin{eqnarray}
\label{E:TWOBODY}
         R_{3D}   & = &  \frac{R_{max}}{2} \bigl( 1 - \cos \chi \bigr)                                                 \\ 
         V_{3D}     & = &  \Biggl[ \frac{2 G M}{R_{max}} \Biggr]^{1/2} \frac{\sin \chi}{1 - \cos \chi}    \\
          t            & = &  \Biggl[ \frac{R_{max}^{\,3}}{8 G M} \Biggr]^{1/2}  \bigl( \chi - \sin \chi \bigr) 
,
\end{eqnarray}
where $R_{3D}$ and $V_{3D}$ are the 3D distance and relative velocity between the two components,
$t$ is the time elapsed since zero separation, $R = 0$, which we assume to be equal to the 
age of the Universe at the redshift of A1750, $t = t_z = 12.34$ Gyr in our concordance \LCDM,
$R_{max}$ is the distance at maximum separation, 
$M = M_C + M_N = 3.8$ \TMSUNFOUR\ is the total mass of the system, 
$G$ is the gravitational constant, and $\chi$ is the development angle.

%
%

The rotation angle between the line of collision
and the plane of the sky, $\theta$, connects the observables, 
the projected distance, $R_p$, and relative radial velocity, $V_r$, 
to these equations: $R_p = R \cos \theta$ and $V_r = V \cos \theta$. Based on our results,
we adopt $R_p = 900$ kpc and $V_r = 960\pm129$ \KMSEC. 
We ignore the error in $R_p$ since is much less than that in $V_r$.
Following \cite{GregoryThompson1984ApJ286}, we substitute the observables into 
Equations~\ref{E:TWOBODY}, 11, and 12, 
and derive a relation between the rotation angle, $\theta$ and the development angle:
\begin{equation} \label{E:THETA}
    \tan \theta =   \frac{t_z \, V_r}{R_p} \frac{ (\cos \chi - 1)^{\,2} }{ \sin \chi \,(\, \chi - \sin \chi \,)}
.
\end{equation}
%
Using Equations~\ref{E:TWOBODY}--\ref{E:THETA}, 
we obtain $V_{3D}$~= 1171$_{-259}^{+231}$ \KMSEC\ and 
$\theta$~= 52.\DEGREE9$_{\;+14.^{\!\!\!\circ}7}^{\;-9.^{\!\!\!\circ}10}$ 
(errors in the angle run from +14.7 to -9.10 because the higher the velocity the
lower the rotation angle). 
The instantaneous relative 3D velocity we obtain from our best fit
model is 1992 \KMSEC\ (with a rotation angle of $\theta$~= 28.\DEGREE8) is significantly higher
than the predictions of the dynamical model signaling a possible tension between 
observations and \LCDM\ predictions for relative velocities of massive clusters. 
However, this comparison is not conclusive because the assumptions of our dynamical model 
are not satisfied. 
Including the mass of the component farther to the South of A1750C and A1750N in the total
mass of the system does not increase the predicted relative velocity of the dynamical 
model significantly since its mass is much less than 10\% of $M_1 + M_2$ 
\citep{HwangLee2009MNRAS397}, and thus does not change our conclusion. 
Note that the predictions for the impact velocity from our simulations are not affected
by the less massive component to the South because we start our simulations 
at the time of impact, and from that point of time the gravitational field is dominated by
the two components we are following.

Summarizing our results: we found that our \FLASH\ simulations can
reproduce the X-ray morphology, projected temperature distribution and
galaxy redshift measurements of A1750 assuming that the 
relative impact velocity is V~= 1460$_{-200}^{+180}$ \KMSEC\ 
and the rotation angle from the plane of the sky is $\theta$~= 28.\DEGREE8$\pm$4.\DEGREE4.
This velocity is higher than the average impact velocity for this mass range, 600 \KMSEC, 
predicted by the concordance \LCDM\ model, but it is within the allowed range.
Our results suggest that we can eliminate the possibility of a large impact velocity 
greater than 2000 \KMSEC\ (at 99.7\% CL) for this merging cluster since that would imply 
a much larger relative radial velocity than the observed.
Note that the quoted uncertainties refer to statistical errors due to observations
based on our idealized cluster merging simulations. 
These errors do not reflect possible systematic errors due to asphericity, substructure 
(visible in the \XMM\ image of the system; see Figure 2 of \citealt{Belset2004AA415}), 
gas falling in along the filament between the two components, and deviations due to
the lack of hydrostatic equilibrium at the outer regions of clusters 
(e.g., \citealt{SerenoET2013MNRAS428,IchikawaET2013ApJ766,ChiuMolnar2012,LaganaET2010,
MolnarET2010ApJ724L1};
and references therein),
and modeling uncertainties in the outer regions of clusters due to weak constraints from
observations, which have not been quantified in this context.
Although the systematic effects might be larger than our statistical errors, 
we do not expect large changes (say a factor of two) in the derived
impact velocity for A1750 from systematic effects because the 
intracluster gas of each component is compressed over a wide radial range,
from $\sim$300 kpc to $\sim$1000 kpc into a 200 kpc region 
from $\sim$300 kpc (the inner limit of the interaction region) 
to $\sim$500 kpc (the position of the contact discontinuity, see Figure~\ref{F:SHOCK}).
Therefore the resulting density and temperature profiles, for comparison with the data, 
should not be very sensitive to the initial conditions of the gas.

\smallskip
\section{Conclusion}
\label{S:Conclusion}

We have been carrying out self--consistent N/body--hydro-dynamical
simulations (including dark matter and gas) to investigate how we can
best determine the impact velocities of merging galaxy clusters.  We
have demonstrated, that the impact velocities can be determined in an
early stage of merging, in the case of A1750, well before the first
core passage with the use of multi--wavelength data and hydrodynamical
simulations. The physical basis which enables us to derive the impact
velocity of a merging system before the first core passage is that the
temperature of the shocked gas depends on the relative velocities of
the two subclusters.  

Observations do not provide the physical temperature of the gas in
clusters directly, but only the projected temperature, which can be
substantially lower than the physical temperature due to the averaging
along the line of sight in projection which could contain a substantial
amount of cold gas. We have shown with our models
that the observed (projected) temperature depends on the angle we
rotate the system towards the observer around the axis perpendicular
to the line connecting the centers of the two subclusters. Even
without rotation, the projected temperature is about 1--2 keV lower
than the physical temperature.  We demonstrated that this degeneracy
can be resolved using the morphology of the merging system and
independently using the observed galaxy redshift information, and
thus a more accurate determination of the impact velocity is possible.
As a consequence, we conclude that the conventional method deriving
the impact velocities for merging systems is not reliable due to two
erroneous assumptions: (1) the measured projected temperature is the
physical temperature; (2) the temperature of the interaction region is
exclusively coming from the shock.

As a case study, we used multi--wavelength observations and numerical
simulations to constrain the impact velocity in the merging galaxy cluster A1750. 
Results from previous weak lensing analysis were used here to constrain the masses
of the two main components, and member galaxy redshifts constrained
the projected LOS instantaneous relative velocities of the two merging clusters. 
We used X--ray measurements to constrain the impact velocity and rotation
angle of the plane of the collision relative to the LOS.  
We found that among our idealized cluster merging simulations with
the input masses constrained to be $M\,^C = 2.0$ and $M\,^N = 1.8$ \TMSUNFOUR, 
the concentration parameters fixed at 8 and 10, and impact parameter at 150 kpc, 
the simulation with a relative velocity of 1460 \KMSEC\ matches all available 
observations well. Impact velocities in excess of 2000 \KMSEC\ are unlikely because, 
either the implied instantaneous relative velocities would exceed the observed LOS 
relative velocity of 900 \KMSEC, or the collision would have to lie very close to the plane 
of the sky, in which case a high velocity impact is excluded by the high projected X--ray 
temperature in the interacting region which is measured to be only about 6 keV by the 
X--ray observations.
Our constraints on the infall velocity are stronger from the radial velocity difference 
derived from optical observations because we do not have strong constraint from 
the gas morphology from the relatively short X--ray exposures.

Impact velocities of merging systems soon after the first core passage 
have been derived from multi-wavelength observations and numerical simulations 
for two clusters 
(the bullet cluster: \citealt{SpriFarr2007MNRAS380p911,MastBurk08MNRAS389p967},
and references therein, and CL0152-1357: \citealt{Molnet2012ApJ748}).
The resulting infall velocities were very high, from 3000 \KMSEC\ to 4800 \KMSEC, 
illustrating the considerable differences between these velocities and infall velocities 
inferred by the \LCDM\ model.
The advantage of focusing on deriving impact velocities of merging
clusters before the first core passage is that merging essentially begins
in this phase and are not yet significantly speeded up, deflected
and suffering considerable ram pressure which is sensitive to the
impact parameter. Several well observed examples of pre--merger
clusters would allow this approach to be applied to existing lensing and X--ray data 
(see \citealt{Mauro2011AA525,Okab2008PASJ60}, and references therein).
The importance of a clear understanding of the relative impact velocities is highlighted 
by the anomalously high velocities inferred for such systems and the difficulty of reproducing 
these velocities within the conventional \LCDM\ framework. Any significant asymmetry in 
the distribution of pre--merger and post--merger relative velocities 
would be of great interest for alternative models of gravity
and for the BEC model of dark matter where coherent non--linear
wavelike behavior of the dark matter \citep{Gonzales2011PhRvD83} may generate
unexpected behavior during interactions.

\acknowledgements

We thank the anonymous referees for valuable comments and suggestions
that improved the presentation of our results, and
K. Umetsu and N. Okabe for discussions about their weak lensing results for A1750.
The code \FLASH\ used in this work was in part developed by the
DOE-supported ASC/Alliance Center for Astrophysical Thermonuclear
Flashes at the University of Chicago.  We thank the Theoretical
Institute for Advanced Research in Astrophysics, Academia Sinica, for
allowing us to use its high performance computer facility for our
simulations.  
This research has made use of the NASA/IPAC
Extragalactic Database (NED) which is operated by the Jet Propulsion
Laboratory, California Institute of Technology, under contract with
the National Aeronautics and Space Administration.

%
%
\bibliographystyle{apj}


\end{document}